\documentclass[aip,jap,superscriptaddress,preprint]{revtex4-1}

\usepackage{graphicx}
\usepackage{epstopdf} 
\usepackage{amssymb} 
\usepackage{amsmath} 

\renewcommand\vec[1]{{\bf #1}}

\begin{document}

\title{Quantum calculations of the carrier mobility in thin films: Methodology, Matthiessen's rule and comparison with semi-classical approaches}

\author{Yann-Michel Niquet}
\email{yniquet@cea.fr}
\author{Viet-Hung Nguyen}
\affiliation{L\_Sim, SP2M, UMR-E CEA/UJF-Grenoble 1, INAC, Grenoble, France}

\author{Fran\c cois Triozon}
\affiliation{CEA, LETI-MINATEC, Grenoble, France}

\author{Ivan Duchemin}
\affiliation{L\_Sim, SP2M, UMR-E CEA/UJF-Grenoble 1, INAC, Grenoble, France}

\author{Olivier Nier}
\author{Denis Rideau}
\affiliation{ST Microelectronics, Crolles, France}

\begin{abstract}
We discuss the calculation of the carrier mobility in silicon films within the quantum Non-Equilibrium Green's Functions (NEGF) framework. We introduce a new method for the extraction of the carrier mobility that is free from contact resistance contamination, and provides accurate mobilities at a reasonable cost, with minimal needs for ensemble averages. We then introduce a new paradigm for the definition of the partial mobility $\mu_{\rm M}$ associated with a given elastic scattering mechanism ``M'', taking phonons (PH) as a reference ($\mu_{\rm M}^{-1}=\mu_{\rm PH+M}^{-1}-\mu_{\rm PH}^{-1}$). We argue that this definition makes better sense in a quantum transport framework as it is free from long range interference effects that can appear in purely ballistic calculations. As a matter of fact, these mobilities satisfy Matthiessen's rule for three mechanisms [surface roughness (SR), remote Coulomb scattering (RCS) and phonons] much better than the usual, single mechanism calculations. We also discuss the problems raised by the long range spatial correlations in the RCS disorder. Finally, we compare semi-classical Kubo-Greenwood (KG) and quantum NEGF calculations. We show that KG and NEGF are in reasonable agreement for phonon and RCS, yet not for SR. We point to possible deficiencies in the treatment of SR scattering in KG, opening the way for further improvements.
\end{abstract}

\maketitle

\section{Introduction}

Device scaling has been a major trend in micro-electronics for almost fifty years, allowing for continuous performance and functionality enhancements. Complementary Metal-Oxide-Semiconductor\cite{Ytterdal} (CMOS) transistors with gate lengths $L_g$ in the 20 nm range are now manufactured at the industrial level thanks to the breakthroughs made in material and device processing. Micro-electronics is now facing new challenges.\cite{ITRS} In particular, extreme scaling calls for innovative device architectures, such as fully-depleted silicon-on-insulator (FDSOI) transistors based on thin films,\cite{Faynot10,Planes12} or multi-gate, short channel nanoscale transistors.\cite{Colinge04,Barraud12} In the sub-10 nm scale, the distinction between material and device modeling is getting increasingly blurred: the electronic and transport properties of the system sharply depart from those of bulk materials and become strongly dependent on the detailed device geometry. Quantum corrections start to be significant with strong sub-band quantization even in the weak inversion regime and source-to-drain tunneling. Confinement also enhances the interactions between, e.g., the silicon channel and the surrounding gate stack, affecting carrier mobilities ever more.\cite{Jin07,Poli09a,Buran09,Poli09b,Poli09c,Persson10,Aldegunde11,Luisier11,Nguyen13,Oh13,Neophytou11,Niquet12a}

Modeling and simulation can play a prominent role in the design and understanding of these devices. The standard simulation toolbox is still, however, mostly based on classical (e.g., drift-diffusion\cite{Granzner06}) and semi-classical simulation methods (Kubo-Greenwood,\cite{Kubo57,Greenwood58} Monte-Carlo\cite{Jacoboni83} or deterministic\cite{Jungemann09} solution of Boltzmann transport equation). This toolbox shall, therefore, be complemented with quantum transport methods, in order to assess the importance of quantum corrections, and in order to work out a multi-scale framework able to address problems at various scales with different levels of approximations. Non-Equilibrium Green's Functions (NEGF) is one of the most versatile approach for that purpose.\cite{Anantram08} In particular, it can deal with quantum confinement, elastic scattering (surface roughness, impurities, ...) and inelastic scattering (phonons) in a seamless way. Although computationally expensive, Green's functions methods have benefited from recent advances in numerical methods and algorithms\cite{Luisier06,Li07,Kazymyrenko08,Cauley11} and from the increasing availability of high performance computing infrastructures.\cite{Luisier08} They can, therefore, be applied to more and more realistic devices, and will certainly take an important place in the design of the ultimate technology nodes.

Although NEGF is primarily intended for large bias (out-of-equilibrium) calculations,\cite{Luisier09,Cavassilas11,Dehdashti13} it can also be used to compute carrier mobilities in a quantum framework. There is already a lot of literature about carrier mobilities within the Green's functions framework, mostly on nanowires.\cite{Poli09a,Buran09,Poli09b,Poli09c,Persson10,Aldegunde11,Luisier11,Nguyen13,Oh13} The carrier mobility is usually extracted from gate length scaling analyses,\cite{Aldegunde11,Luisier11,Nguyen13} or from a separation between the ballistic\cite{Datta97,Shur02} and diffusive components of the current.\cite{Buran09,Poli09b,Poli09c,Persson10} In this work, we discuss the strengths and weaknesses of these approaches, and propose a new method that gives very accurate results at a reasonable cost (Sections \ref{sectionDevices} and \ref{sectionMethodo}). We apply this methodology to thin silicon films (thickness in the 2-10 nm range), on devices with unprecedented size (length up to 95 nm and width up to 30 nm). We introduce a new paradigm for the definition of the partial mobility $\mu_{\rm M}$ associated with a single elastic mechanism ``M'', taking phonons (PH) as a reference ($\mu_{\rm M}^{-1}=\mu_{\rm PH+M}^{-1}-\mu_{\rm PH}^{-1}$). We argue that this definition makes better sense in a quantum transport framework, as it mitigates long range interference effects that can appear in ballistic (no phonons) calculations. We also discuss the problems raised by long range spatial correlations appearing in, e.g., remote Coulomb scattering (RCS, section \ref{sectionRCS}) We show that our partial mobilities actually satisfy Matthiessen's rule much better than the usual, single mechanism calculations (Section \ref{sectionMatthiessen}). Finally, we compare NEGF mobilities with KG calculations (Section \ref{sectionComparisons}). We show that NEGF and KG agree reasonably well for phonons and RCS scattering, yet not for SR. We discuss possible reasons for these discrepancies.

\section{Devices and methodologies}
\label{sectionDevices}

\begin{figure}
\includegraphics[width=.50\columnwidth]{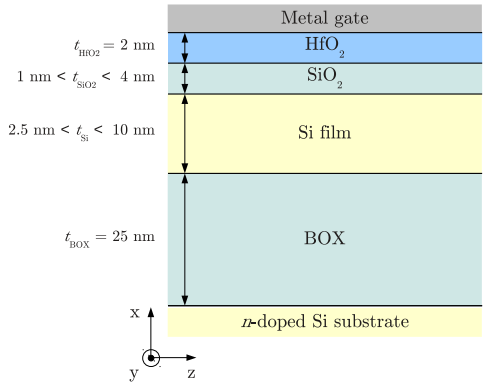} 
\caption{Schematics of the simulated FDSOI devices.\label{FigDev}}
\end{figure}

The devices considered in this work (Fig. \ref{FigDev}) are undoped, $(100)$ FDSOI films with thickness ranging from $t_{\rm Si}=2.5$ nm to $t_{\rm Si}=10$ nm. The buried oxide (BOX) below the film is 25 nm thick, and the $n$-doped ($N_d=10^{18}$ cm$^{-3}$) silicon substrate acting as a back gate is grounded. The front gate (FOX) stack is made up of a layer of SiO$_2$ (with thickness $1<t_{\rm SiO_2}<4$ nm and dielectric constant $\varepsilon=3.9$), and of a layer of HfO$_2$ (with thickness $t_{\rm HfO_2}=2$ nm and dielectric constant $\varepsilon=22$). 

The current is computed in a self-consistent NEGF framework,\cite{Anantram08} on top of the effective mass approximation (EMA).\cite{Bastard} The longitudinal mass is $m_l^*=0.916\ m_0$ and the transverse mass $m_t^*=0.191\ m_0$ in silicon. The effective mass in SiO$_2$ is $m^*=0.5\ m_0$, and the Si/SiO$_2$ barrier is 3.15 eV high. The NEGF equations are solved on a finite differences grid, in a fully coupled mode space approach.\cite{Wang04} Details can be found in Appendix \ref{AppendixNEGF}.

NEGF can deal with phonons,\cite{Jin06} surface roughness\cite{Buran09} (SR) and remote Coulomb scattering\cite{Poli09c} (RCS) in a seamless way. We account for intra-valley acoustic phonon scattering (deformation potential\cite{Esseni03a} $D_{\rm ac}=14.6$ eV), and for inter-valleys scattering by the 3 $f$-type and 3 $g$-type processes of Ref.~\onlinecite{Jacoboni83}.
 
Random SR profiles are generated as in Ref.~\onlinecite{Buran09}, with an exponential auto-covariance function:\cite{Goodnick85}
\begin{equation}
F_{\rm SR}(\vec{r})=\langle\delta h(\vec{R})\delta h(\vec{R}+\vec{r})\rangle=\Delta^2 e^{-\sqrt{2}r/\ell_c}\,,
\label{eqSR}
\end{equation}
where $\delta h(\vec{r})$ is the variation of the surface height at point $\vec{r}$. The typical rms is $\Delta=0.47$ nm and the correlation length $\ell_c=1.3$ nm (same parameters as in Ref. \onlinecite{Jin09}). The FOX Si/SiO$_2$, SiO$_2$/HfO$_2$ and HfO$_2$/gate interfaces are conformal, but the FOX and BOX interfaces are uncorrelated.

As for RCS, we generate random distributions of charges at the SiO$_2$/HfO$_2$ interface (with a given density $n_{\rm RCS}$), then solve Poisson's equation for the RCS potential. The caveats of this solution will be discussed in section \ref{sectionRCS}.

The semi-classical Kubo-Greenwood calculations discussed in this work have been performed with the commercial ``Sentaurus Device'' solver of Synopsys.\cite{sband}

\section{Extracting mobilities from NEGF calculations}
\label{sectionMethodo}

\subsection{Definitions and problems}
\label{subsectionDef}

In general, the low-field resistance of a channel with length $L$ can be written:
\begin{equation}
R(L)=\frac{V}{I}=R_c+R_0+\frac{L}{n_{\rm 1d}\mu e}\,,
\label{eqRl}
\end{equation}
where $V$ is the (small) drain-source bias, $I$ is the current, $n_{\rm 1d}$ is the carrier density per unit length, $\mu$ is the carrier mobility and $e$ is the electron charge. The first term, $R_c$, is a ``contact'' resistance accounting for backscattering in the access areas and/or at the interface between the access areas and the channel. It is, therefore, extrinsic to the channel. The second term is the so-called ``ballistic'' resistance of the channel,\cite{Datta97,Shur02} while the third, $\propto L$ term is the classical ``diffusive'' resistance. At zero temperature, the resistance of a purely ballistic channel, $R_0=1/(G_0N_{\rm m})=12.9{\rm k}\Omega/N_{\rm m}$, is limited by the number of modes (1D sub-bands) $N_m$ carrying current. At finite temperature $T$, assuming Maxwell-Boltzmann statistics and a single transport mass $m^*$ for all sub-bands:\cite{Nguyen13}
\begin{equation}
\frac{1}{R_0}=-\frac{2e^2}{h}\int dE\left(\frac{\partial f}{\partial E}\right)t(E)=\frac{n_{\rm 1d}e^2}{\sqrt{2\pi m^*kT}}\,,
\label{eqR0}
\end{equation}
where $f(E)=\exp[(E-\mu)/kT]$ is the distribution function, and $t(E)$ the transmission function, which is equal to the number of 1D sub-bands at energy $E$. Although the above assumptions may not hold in general, Eq. (\ref{eqR0}) nicely illustrates the main trends followed by the ballistic resistance. In particular, both the diffusive and ballistic resistance decrease with the width $W$ of 2D devices, since $n_{\rm 1d}=n_{\rm 2d}W$, where $n_{\rm 2d}$ is the sheet density in the channel.

We emphasize, at this point, that Eq. (\ref{eqRl}) is only valid in ``long enough'' channels. $L$ shall not only be larger than the mean free path $\ell_e$ to reach the diffusive regime; It must also be much larger than the typical correlation length $\ell_c$ of the disorder so that the carriers sample a representative set of configurations along their way from source to drain (``self-averaging''). Variability around Eq. (\ref{eqRl}) increases with decreasing $L$, and the mobility must primarily be understood as a long channel concept, or as an average figure for channels shorter than a few $\ell_c$'s. As discussed in paragraph \ref{sectionRCS}, $\ell_c$ can range from a $1-2$ nanometers (e.g., SR) to $\approx 10$ nanometers (e.g., RCS).

In principle, the carrier mobility can be computed in different ways with a real space NEGF code. The ballistic resistance $R_0$ can be obtained independently from a purely ``ballistic'' calculation (no scattering), and the mobility extracted from the data for a single length $L$ using Eq.~(\ref{eqRl}).\cite{Buran09,Poli09b,Poli09c,Persson10} This, however, presumes that the contact resistance $R_c$ is negligible, and that the length $L$ of the channel is perfectly well defined (where does the channel really start and end in the device ?). Alternatively, the current can be computed in channels with various lengths $L$, then $R_c+R_0$ and $\mu$ fitted to the $R(L)$ data (gate length scaling analysis).\cite{Aldegunde11,Luisier11,Nguyen13} This method is {\it a priori} immune to contact resistance contamination and to channel length misestimates. Indeed, a systematic error $\Delta L$ on the channel length will not change the slope of the $R(L)$ data (hence the mobility), but only the apparent contact and ballistic resistance $R_c+R_0$. Yet the $R(L)$ data can be very noisy, since different $L$ usually correspond to different realizations of the disorder (with different diffusive resistances but also possibly different $R_c$'s). Therefore, accurate mobilities practically call for averages over large numbers of samples, especially if $L$ can not be made very long.

Also, as shown by Eqs.~(\ref{eqRl}) and (\ref{eqR0}), the resistance of the channel has a prevalent $1/n_{\rm 1d}$ behavior. It is, therefore, essential to compare data computed at the same carrier density (whatever the method). This raises at once the question ``What is the carrier density in the channel ?'', which is not trivial. In a disordered (e.g., rough) channel, the carrier density is intrinsically non uniform, hence not univocally defined.

We therefore need to design a method that is free from the above limitations to the largest possible extent, i.e. {\it i}) free from contact resistance contamination; {\it ii}) free from channel length misestimates; {\it iii}) with minimal needs for ensemble averages, and {\it iv}) with a well defined prescription for the density.

\subsection{Methodology and example}
\label{subsectionSR}

We have set-up the following methodology to deal with the above issues: We prepare a sample of disorder (e.g., surface roughness) with periodic boundary conditions over length $L_s$ and width $W_s$. We then build devices with lengths $L(N)=2L_c+NL_s$ made of this sample repeated $N$ times and connected to access areas with length $L_c$ on both source and drain sides. The resistance of these devices is therefore expected to follow an arithmetic progression $R(N)=R_c+R_0+NR_s$, representative of a series of $N$ segments with ballistic resistance $R_0$ and diffusive resistance $R_s$, connected to access areas with total resistance $R_c$. The noise on the $R(N)$ data shall be minimal since all segments are identical, allowing for an accurate extraction of the sample resistance $R_s$ and mobility $\mu=L_s/(n_{\rm 1d}eR_s)$ according to Eq. (\ref{eqRl}). In particular, this method is completely free from contact resistance contamination as $R_c$ shall be the same whatever $N$, all devices showing, by design, exactly the same interfaces between source/drain and channel. It is also immune to channel length misestimates, as the period $L_s$ of the disorder is perfectly well defined. This shall decrease the number of samples needed to converge ensemble averages. 

\begin{figure}
\includegraphics[width=.50\columnwidth]{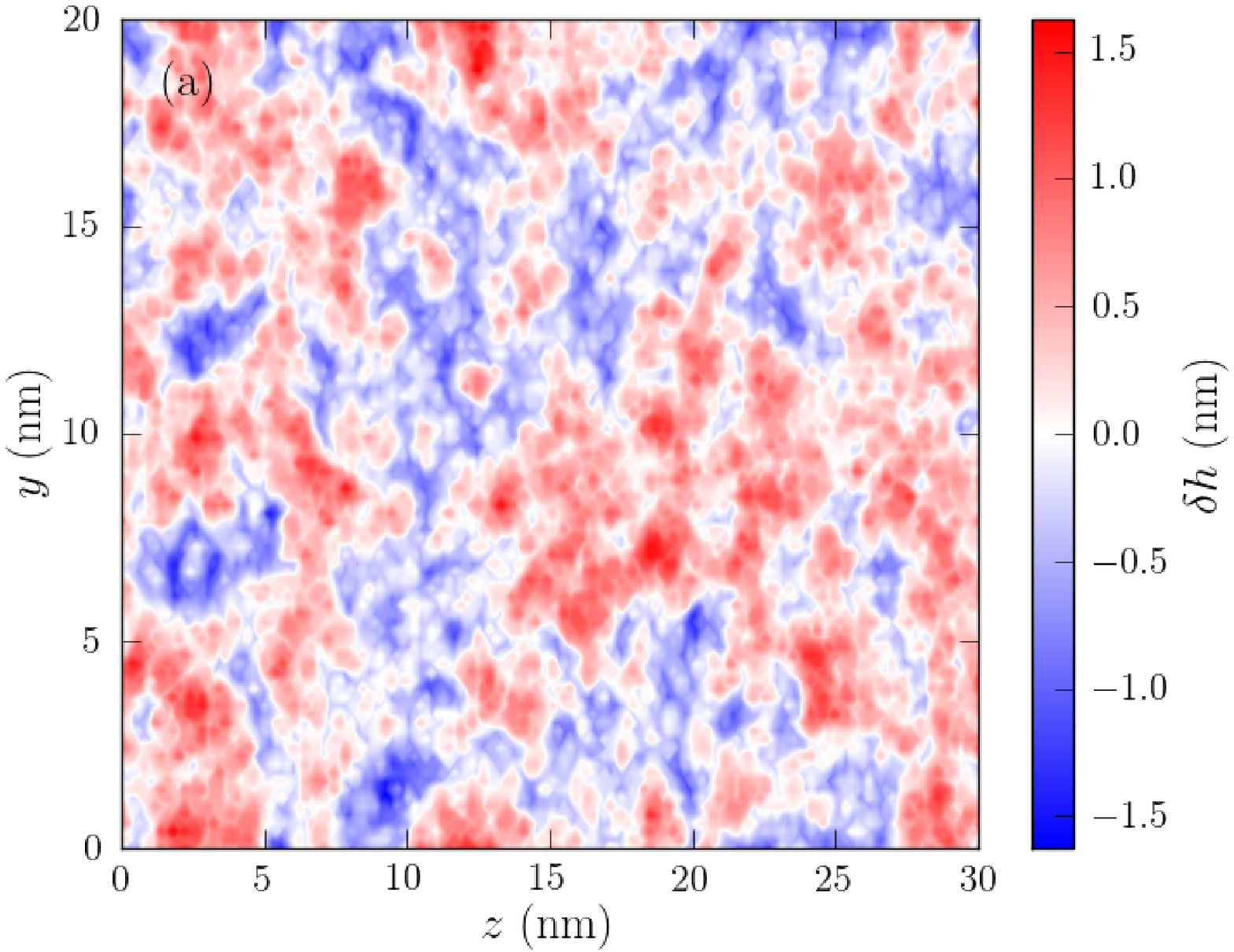} 
\includegraphics[width=.50\columnwidth]{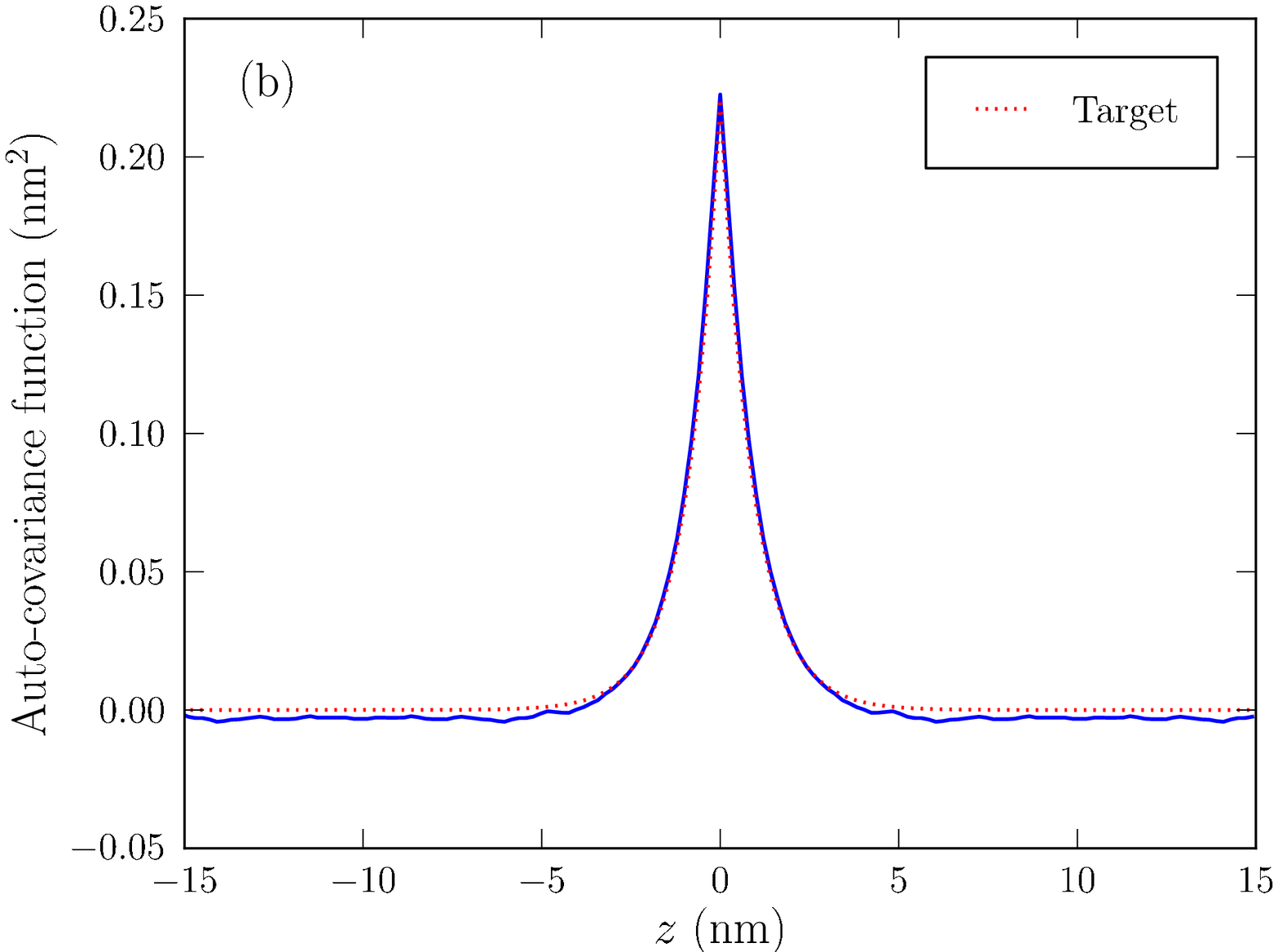}
\caption{(a) A periodic $W_s=20{\rm\ nm}\times L_s=30{\rm\ nm}$ sample of surface roughness. (b) The spatial (solid blue line) and target (dotted red line) auto-correlation functions of this sample, computed along the $z$ (transport) axis.\label{FigSRprofile}}
\end{figure}

For the sake of illustration, we focus on phonons+surface roughness (PH+SR) scattering in thin film devices -- though the methodology is valid in 1D trigate or nanowire devices as well. A typical, periodic $W_s=20{\rm\ nm}\times L_s=30{\rm\ nm}$ sample of SR is shown in Fig.~\ref{FigSRprofile}, along with the target [Eq. (\ref{eqSR})] and calculated spatial auto-correlation functions:
\begin{equation}
F(\vec{r})=\frac{1}{S}\int_{\cal S} d^2\vec{R}\,\delta h(\vec{R})\delta h(\vec{R}+\vec{r})\,,
\label{eqSR2}
\end{equation}
where ${\cal S}$ is the Si/SiO$_2$ interface with surface $S$. We will come back in section \ref{sectionRCS} to the importance of this function.

\begin{figure}
\includegraphics[width=.50\columnwidth]{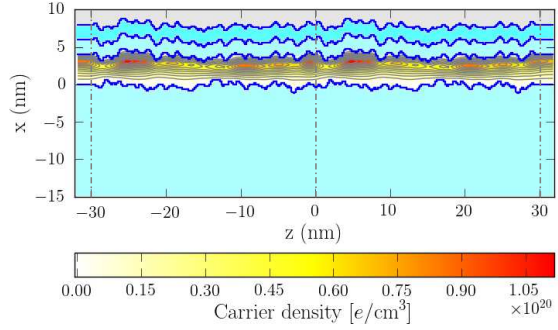} 
\caption{Carrier density in the $N=2$ ($L=63.6$ nm long) device at gate voltage $V_{\rm gs}=1.6$ V. From bottom to top, the BOX, Si film, SiO$_2$, HfO$_2$ and gate layers are clearly visible, as well as the 30 nm long SR sample, repeated twice. [$t_{\rm Si}=4$ nm, $t_{\rm SiO_2}=2$ nm]\label{FigDevice}}
\end{figure}

We build devices made of this sample repeated once [$L(1)=33.6$ nm], twice [$L(2)=63.6$ nm, see Fig.~\ref{FigDevice}] and up to three times [$L(3)=93.6$ nm]. There is a $L_c=1.8$ nm long, undisordered ``contact'' region on each side of the device, which has no sizable influence on the extracted $R_s$ (though it has on $R_c$). Note that the devices are fully gated from left to right. The density of carriers is thus controlled by the electrochemical potential $\mu_s$ in the source, the doping density and the gate voltages. $\mu_s$ was set as the electrochemical potential in a (remote) source with $n$-type doping $N_d=5\times10^{18}$ cm$^{-3}$. The choice of $\mu_s$ is pretty irrelevant in single gate devices (this merely shifts the $I(V_{\rm gs})$ characteristics along the front gate $V_{\rm gs}$ axis), but can have some impact on the mobility in double gate devices, due to the interplay between the front and back gate electric fields.

As for the definition of $n_{\rm 1d}$, the average density in the device practically yields the best (most linear) fits to the $R(N)$ data. We therefore stick to this definition, which makes sense as the variations around the average density can be interpreted as the response (screening) of the carriers to the disorder. Note that $n_{\rm 1d}$ is very little sensitive to the the length $L_c$ of the contact regions, since there is no junction between a highly doped source/drain and the channel in the simulation box.

We sweep the gate voltage from $V_{\rm gs}=0$ V to $V_{\rm gs}=1.8$ V and monitor the current, average density and effective electric field in the devices. The drain-source voltage is $V_{\rm ds}(1)=2$ mV, $V_{\rm ds}(2)=4$ mV, and $V_{\rm ds}(3)=6$ mV, in order to extract the mobility at low, but constant longitudinal electric field. The average density and effective electric field are weakly dependent on the device length $L$ (within $\pm 1\%$); However, as discussed in section \ref{subsectionDef}, the current has a strong dependence on $n_{\rm 1d}$, so that it is best to compare resistances computed at the same density. Therefore, we first fit $n_{\rm 1d}R(n_{\rm 1d})$ with a spline (for each $N$), then interpolate $R(n_{\rm 1d})$ on a grid of target densities. This slightly improves the quality of the linear regression on the $R(N)$ data.

\begin{figure}
\includegraphics[width=.50\columnwidth]{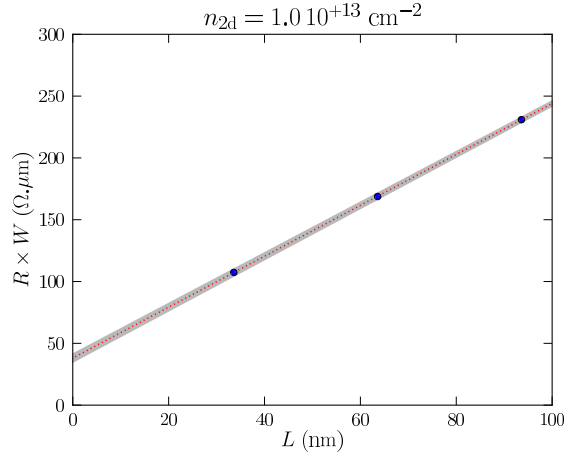} 
\caption{Phonons+SR resistance (times device width) as a function of length for a sample like Fig. ~\ref{FigSRprofile}. The dotted red line is a linear regression with Eq.~(\ref{eqRl}), which yields $(R_c+R_0)W=38.04\Omega.\mu{\rm m}$ and $\mu_{\rm PH+SR}=303$ cm$^2$/V/s (carrier density $n_{\rm 2d}=10^{13}$ cm$^{-2}$). The shaded gray area is the 95\% confidence interval. [$t_{\rm Si}=4$ nm, $t_{\rm SiO_2}=2$ nm]\label{FigR2R3R}}
\end{figure}

\begin{figure}
\includegraphics[width=.50\columnwidth]{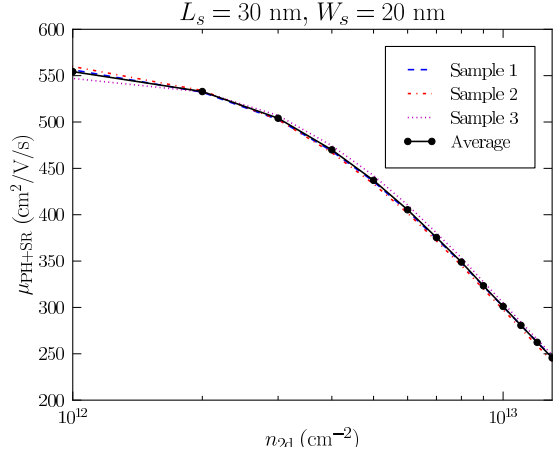} 
\caption{The phonon+SR limited mobility as a function of carrier density, for three different SR samples with size $W_s=20{\rm\ nm}\times L_s=30{\rm\ nm}$, along with the average mobility $\mu_{\rm avg}^{-1}=(\mu_1^{-1}+\mu_2^{-1}+\mu_3^{-1})/3$. [$t_{\rm Si}=4$ nm, $t_{\rm SiO_2}=2$ nm]\label{FigvariPHSR}}
\end{figure}

As an example, the $R(N)$ data computed for a specific SR sample are plotted in Fig.~\ref{FigR2R3R} (carrier density $n_{\rm 2d}=10^{13}$ cm$^{-2}$). As expected, the data lie on a straight line, allowing for an unambiguous fit $(R_c+R_0)W=38.04\ \Omega.\mu{\rm m}$ and $\mu_{\rm PH+SR}=303$ cm$^2$/V/s ($\pm 2.25\%$ with $95\%$ confidence). For practical purposes, we most often compute the $N=1$ and $N=2$ devices only, and check the $N=3$ data for a few, critical cases. The calculated phonons+SR mobility is plotted as a function of carrier density in Fig.~\ref{FigvariPHSR}, for $N_s=3$ different SR samples, along with the average:
\begin{equation}
\mu_{\rm avg}^{-1}=\frac{1}{N_s}\sum_{i=1}^{N_s} \mu_i^{-1}\,,
\label{eqmuavg}
\end{equation}
where $\mu_i$ is the mobility extracted from the $i^{\rm th}$ sample. There is little variability -- hence practically no need for ensemble averages, as the samples are longer than the mean free path and much longer than the correlation length ($\ell_c=1.3$ nm) of the disorder. Using the relations $\mu=e\tau/m^*$, where $\tau$ is the average scattering time, and $\ell_e=v\tau$, where $v$ is the thermal velocity $\frac{1}{2}m^*v^2=kT$, we indeed estimate mean free paths in the $2-19$ nm range for mobilities $100\le\mu\le 800$ cm$^2$/V/s. Convergence with respect to the sample size will be discussed in more detail in the next subsection.

The ballistic resistance $R_0W=34.44\ \Omega.\mu{\rm m}$ of the the same device has also been computed independently (switching off all scattering mechanisms). It is, as expected, significantly lower than the $R(L=0)W=38.04\ \Omega.\mu{\rm m}$ extrapolation of the NEGF data. This is due, primarily, to the resistance $R_c$ associated with the two $L_c=1.8$ nm long contacts on both sides of the disordered channel, and with the backscattering at the contact/channel interfaces. This might also be due to the mismatch between the ballistic resistance of the purely ballistic channel and the ``ballistic'' resistance of the rough channel, which will be limited by the thinnest parts of the device, and is therefore expected to be slightly larger. 

\begin{figure}
\includegraphics[width=.50\columnwidth]{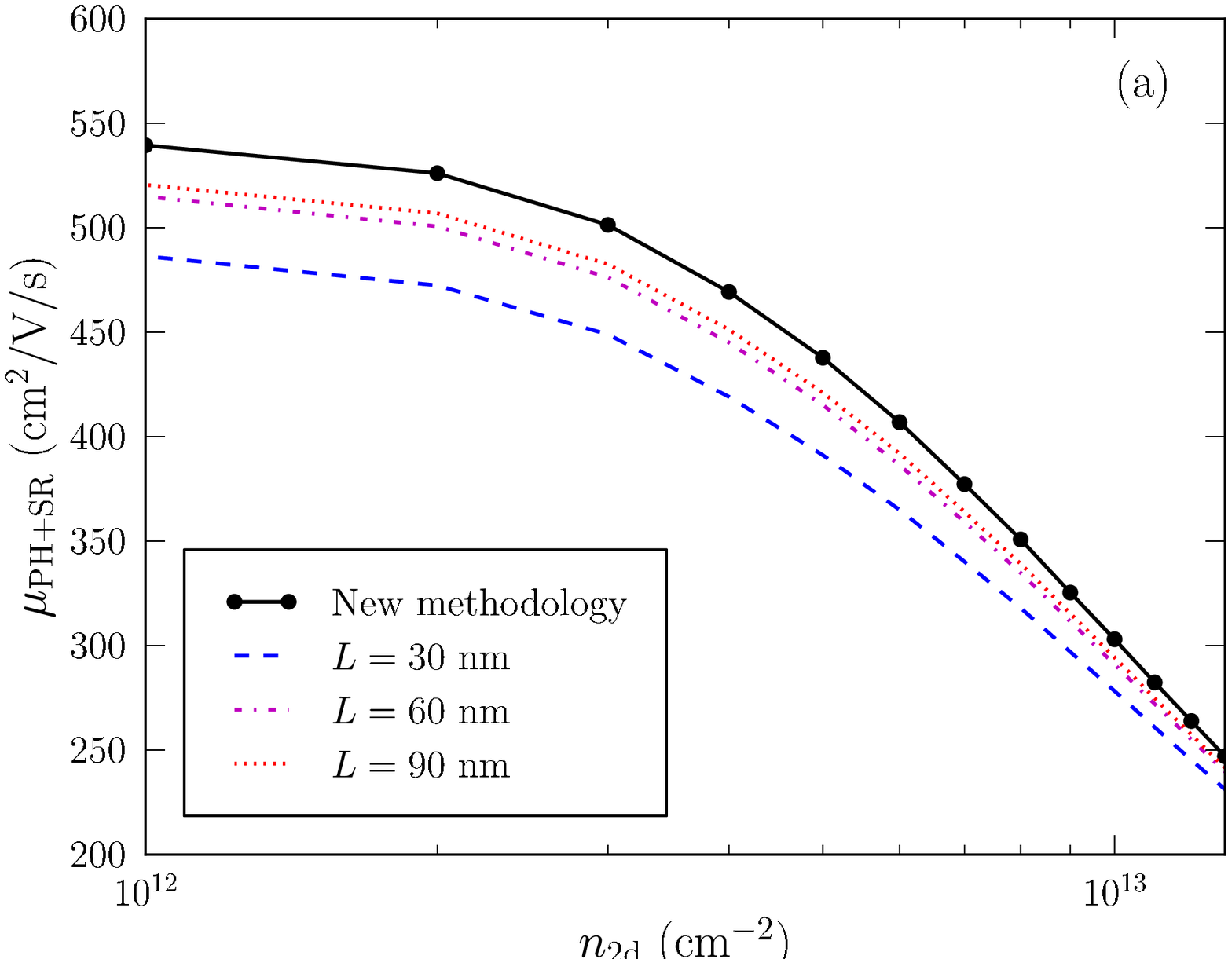} 
\includegraphics[width=.50\columnwidth]{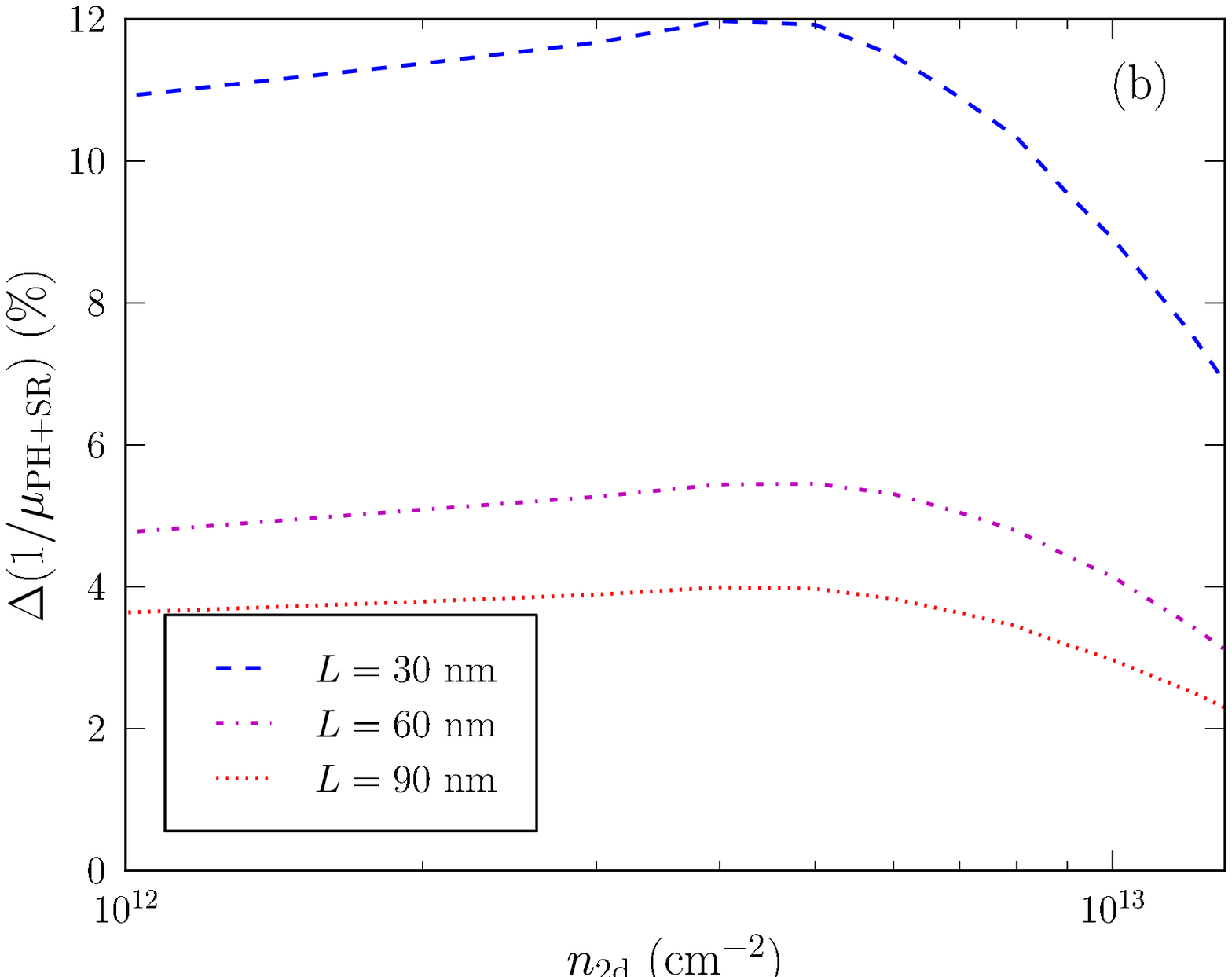} 
\caption{(a) The phonon+SR limited mobility extracted from our methodology and from single length calculations with different $L$'s [Eq.~(\ref{eqmuball})], and (b) the relative  difference between our methodology and the single length calculations. [$t_{\rm Si}=4$ nm, $t_{\rm SiO_2}=2$ nm]\label{FigScaling}}
\end{figure}

Since the above methodology requires at least two calculations on ``long'' devices, it is worth comparing the results with mobilities extracted from (cheaper) single length data as:
\begin{equation}
\mu(L)=\frac{L}{n_{\rm 2d}e[R(L)-R_c-R_0]W}\,.
\label{eqmuball}
\end{equation}
Using the data of Fig.~\ref{FigR2R3R} with the extrapolated $(R_c+R_0)W=38.04\Omega.\mu{\rm m}$ would of course yield exactly the same mobility $\mu_{\rm PH+SR}=303$ cm$^2$/V/s whatever $L$ but presumes that we have computed at least two long devices already. Instead, we use our ``best'' simple estimate for $R_c+R_0$, that is the phonon limited resistance of a very short 3.6 nm long channel (mimicking the two access areas on the source and drain sides). Accordingly, we use the length $L=NL_s$ of the channel in Eq. (\ref{eqmuball}). The mobilities extracted for different $L$'s are compared with our methodology on Fig.~\ref{FigScaling}. As expected from Eq. (\ref{eqmuball}), any error on $R_c+R_0$ results in a $\propto 1/L$ correction on $1/\mu$, leading to an apparent length dependent mobility.\cite{Poli09b} In that particular case, the error is as large as $12\%$ on the 30 nm long device, and is still $\approx 4\%$ on the 90 nm long device. The error would be up to $25\%$ in the 30 nm long device if we had used the ballistic resistance $R_0W=34.44\ \Omega.\mu{\rm m}$ as an approximation for $R_c+R_0$. We therefore conclude that our methodology yields the best balance between accuracy and efficiency for NEGF mobility calculations.

\subsection{Phonons as a reference}
\label{subsectionPhonons}

The above methodology yields the phonons+SR limited mobility $\mu_{\rm PH+SR}$. What about the phonons and SR limited mobilities ?

The phonon limited mobility $\mu_{\rm PH}$ can easily be extracted in the same way. The phonon limited resistance is, actually, strictly proportional to $L$, since the electron-phonon interaction is ``intrinsic'' to the material, at variance with the ``extrinsic'' disorders such as SR and RCS.\cite{Luisier11,Aldegunde11} We might, arguably, compute the SR limited mobility along the same lines (simply switching off phonons in the above calculation). There are, however, two difficulties with this procedure. The first one is computational: Calculations are actually faster with than without phonons. Of course, one has to achieve both Poisson and ``Born'' (self-energies) self-consistency in NEGF calculations with phonons, which costs extra iterations (see Appendix \ref{AppendixNEGF}). On the other hand, inelastic scattering smooths Van Hove singularities in the spectral functions (local density of states, etc...), so that the number of energy points needed to integrate these spectral functions [e.g., Eq.~(\ref{eqQG})] can be up to $10\times$ smaller than in a ``ballistic'' (no phonons) calculation. Hence, with an efficient preconditioning scheme for the self-consistent loops, the computational cost of a calculation with phonons can be much lower than the one of a ballistic calculation. The second issue is about physics. Phonons do break phase coherence and mitigate localization and other long range interference effects.\cite{Kramer93} NEGF (or other quantum) calculations without phonons do not, therefore, necessarily give an accurate picture of the physics of the devices at room temperature. This is especially sensitive in 1D devices such as nanowires, but can also affect 2D devices.

As hinted above, the electron-phonon interaction plays a particular role in the physics of the devices. It is the only inelastic and ``intrinsic'' scattering mechanism. Its interactions with elastic mechanisms (leading, in particular, to decoherence) suggest that we may choose it as a reference frame in quantum calculations. We therefore define an effective SR limited mobility from Matthiessen's rule:
\begin{equation}
\mu_{\rm SR, eff}^{-1}=\mu_{\rm PH+SR}^{-1}-\mu_{\rm PH}^{-1}\,,
\label{eqmueff}
\end{equation}
where $\mu_{\rm PH}$ is the phonon limited mobility and $\mu_{\rm PH+SR}$ the phonon+SR limited mobility, both computed with the above NEGF methodology.

\begin{figure}
\includegraphics[width=.50\columnwidth]{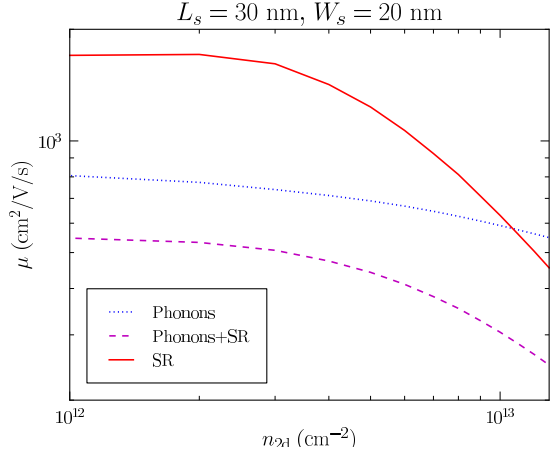} 
\caption{The phonon limited and phonon+SR limited mobilities computed with NEGF, and the effective SR limited mobility obtained from Eq.~(\ref{eqmueff}). [$t_{\rm Si}=4$ nm, $t_{\rm SiO_2}=2$ nm]\label{FigmuPHSR}}
\end{figure}

We stress that $\mu_{\rm PH}$ and $\mu_{\rm PH+SR}$ are unambiguously given by NEGF. Although the above $\mu_{\rm SR, eff}$ might differ from a direct calculation, the combination of $\mu_{\rm SR, eff}$ with the phonon limited mobility $\mu_{\rm PH}$ yields the ``exact'' total mobility $\mu_{\rm PH+SR}$, which is the only important figure (if there are no other scattering mechanisms such as Coulomb traps). The validity of Matthiessen's rule has actually been much debated in the literature.\cite{Walukiewicz79,Stern80,Takeda81,Saxena85,Fischetti02,Esseni11,Chen13} Yet, as shown in section \ref{sectionMatthiessen}, $\mu_{\rm PH}$ and the {\it effective} mobilities $\mu_{\rm SR, eff}$ and $\mu_{\rm RCS, eff}$ defined with respect to phonons can be combined with very good accuracy using Matthiessen's rule. This makes a clear case for Eq. (\ref{eqmueff}) as a definition of the single mechanism mobility.

\begin{figure}
\includegraphics[width=.50\columnwidth]{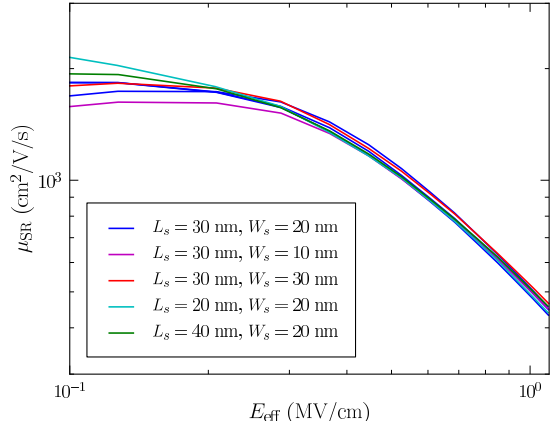} 
\caption{The SR limited mobility as a function of effective field, computed in samples with different lengths $L_s$ and widths $W_s$. [$t_{\rm Si}=4$ nm, $t_{\rm SiO_2}=2$ nm]\label{FigvariSR}}
\end{figure}

As for surface roughness, the effective mobility obtained from Eq.~(\ref{eqmueff}) is plotted in Fig. \ref{FigmuPHSR} for a particular SR sample. The SR limited mobility follows the expected trend as a function of carrier density,\cite{Esseni03a,Jin09} with a plateau at small density (limited by both FOX and BOX SR), and a fast decrease at large density (due to strong inversion at the FOX interface). The mobilities computed on other samples with the same or different sizes are shown in Fig. ~\ref{FigvariSR}. As expected from Fig.~\ref{FigvariPHSR}, there is little sample to sample variability, except possibly at low density/effective field where SR is not the dominant mechanism. The SR limited mobility is converged in $W_s=20{\rm\ nm}\times L_s=30{\rm\ nm}$ samples.

\begin{figure}
\includegraphics[width=.50\columnwidth]{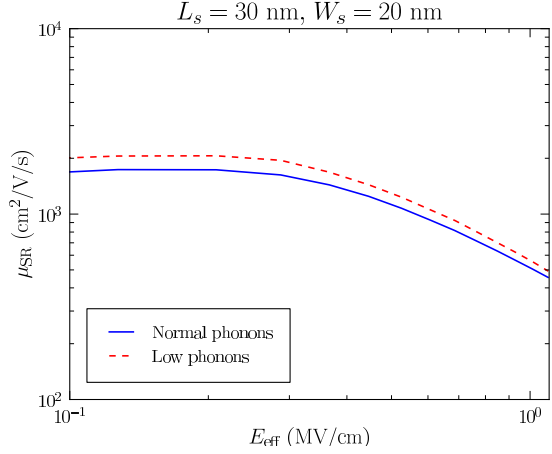} 
\caption{The SR limited mobility extracted from Eq.~(\ref{eqmueff}) for ``normal'' and ``low'' ($\sqrt{2}\times$ weaker deformation potentials) phonons. [$t_{\rm Si}=4$ nm, $t_{\rm SiO_2}=2$ nm]\label{FiglowPH}}
\end{figure}

To demonstrate that Eq. (\ref{eqmueff}) is meaningful, we have recomputed the effective SR mobility using deformation potentials $\sqrt{2}$ times weaker (that is a phonon limited mobility two times larger). As shown in Fig. \ref{FiglowPH}, the resulting $\mu_{\rm SR}$ is little affected ($<10\%$), showing that the methodology is robust. We will further discuss the applicability of Matthiessen's rule in section \ref{sectionMatthiessen}.

\section{The importance of the auto-correlation functions: The case of Remote Coulomb Scattering}
\label{sectionRCS}

In this section, we discuss the problems arising when the samples can not be much longer than the correlation length of the disorder, and possible workarounds. 

Remote Coulomb Scattering (RCS) is the scattering of carriers in the channel by remote charges at the SiO$_2$/HfO$_2$ interface.\cite{Gamiz03,Esseni03b,Casse06,Barraud08,Toniutti12} It is believed to be a major limiting mechanism in high-$\kappa$ gate stacks. It is modeled in Kubo-Greenwood solvers as the scattering by independent, uncorrelated point charges at the interface. It is, therefore, tempting (and straightforward) to mimic the same assumptions in NEGF by randomly distributing test charges at the interface between SiO$_2$ and HfO$_2$, then solve Poisson's equation for the scattering potential $V_{\rm test}(\vec{r})$. As in the case of SR, we generate periodic samples of RCS disorder in $W_s=20{\rm\ nm}\times L_s=30{\rm\ nm}$ supercells.

\begin{figure}
\includegraphics[width=.50\columnwidth]{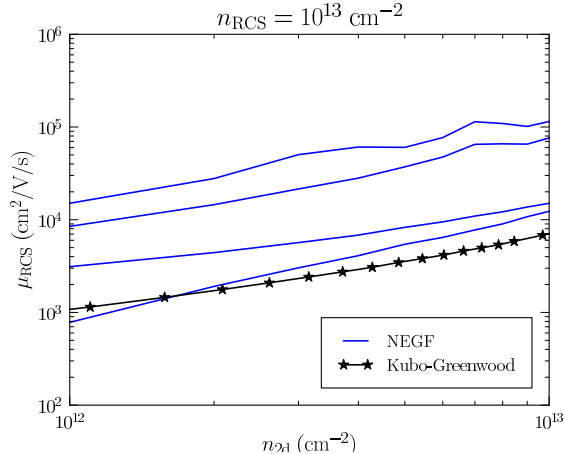} 
\caption{The RCS limited mobility as a function of carrier density, computed for different distributions of charges at the SiO$_2$/HfO$_2$ interface (positive charges with density $n_{\rm RCS}=10^{13}$ cm$^{-2}$). [$t_{\rm Si}=4$ nm, $t_{\rm SiO_2}=1$ nm]\label{FigvariRCS1}}
\end{figure}

The effective RCS mobilities extracted on different samples are plotted as a function of the carrier density in Fig.~\ref{FigvariRCS1}. In these devices, the interfaces are smooth, the SiO$_2$ FOX oxide is only 1 nm thick (to enhance RCS) and the density of charges at the SiO$_2$/HfO$_2$ interface is $n_{\rm RCS}=10^{13}$ cm$^{-2}$ (60 positive charges in the sample). The results of a Kubo-Greenwood calculation have been added for comparison [the same methodology, Eq. (\ref{eqmueff}), was used to extract the mobility for consistency]. Although the trends are well reproduced by NEGF, the sample to sample variability is impressive ($>$ one order of magnitude). These data are clearly representative of the large, Coulomb-induced variability expected in short channel devices,\cite{Asenov09} yet they do not give any clue about the long channel mobility.

These deficiencies can be understood from second-order perturbation theory (which underlies Kubo-Greenwood calculations and is the leading order of NEGF calculations). For the sake of simplicity, we discard electron-phonon interactions here. In the presence of a scalar random potential potential $V(\vec{r})$, the Green's function of the film, $G(E)$, can be expanded to second order in $V$ and the unperturbed Green's function $G_0(E)$ as:\cite{Mahan,Oh13}
\begin{equation}
G(E)=G_0(E)+G_0(E)VG_0(E)+G_0(E)VG_0(E)VG_0(E)+...\,.
\label{eqGDyson}
\end{equation}
The ensemble averaged Green's function $\langle G(E)\rangle$ therefore reads:
\begin{equation}
\langle G(E)\rangle=G_0(E)+G_0(E)\langle V\rangle G_0(E)+G_0(E)\langle VG_0(E)V\rangle G_0(E)+...\,,
\end{equation}
where $\langle...\rangle$ is an average over different samples. Including $\langle V\rangle(\vec{r})$ into the unperturbed Hamiltonian $H_0$ and Green's function $G_0=[E-H_0]^{-1}$, we might as well write:
\begin{equation}
\langle G(E)\rangle=G_0(E)+G_0(E)\langle \delta VG_0(E)\delta V\rangle G_0(E)+...\,,
\label{eqDyson}
\end{equation}
where $\delta V(\vec{r})=V(\vec{r})-\langle V\rangle(\vec{r})$. The above equation can be cast in the form:
\begin{equation}
\langle G(E)\rangle=G_0(E)+G_0(E)\Sigma(E)G_0(E)+...\,,
\label{eqG2ndorder}
\end{equation}
where:
\begin{equation}
\Sigma(\vec{r}, \vec{r}^\prime, E)=G_0(\vec{r}, \vec{r}^\prime, E)\left\langle\delta V(\vec{r})\delta V(\vec{r}^\prime)\right\rangle
\label{eqSRS}
\end{equation}
is the self-energy associated with the random potential $V(\vec{r})$. The above equations are nothing else than a non self-consistent, yet conserving Born approximation.\cite{Mera12} 

At that level, the disorder can be completely characterized by the ensemble averaged auto-covariance function of the potential:
\begin{equation}
F(\vec{r}, \vec{r}^\prime)=\langle\delta V(\vec{r})\delta V(\vec{r}^\prime)\rangle\,.
\end{equation}
In the case of RCS,
\begin{equation}
V(\vec{r})=\sum_{i=1}^N \nu(\vec{r}-\vec{R}_i)
\end{equation}
where $\nu(\vec{r}-\vec{R}_i)$ is the potential created by a single charge at position $\vec{R}_i$, and $N$ is the number of RCS charges in the sample. Assuming uncorrelated charge positions, 
\begin{eqnarray}
F(\vec{r}, \vec{r}^\prime)&=&\sum_{i,j=1}^N\left\langle\delta\nu(\vec{r}-\vec{R}_i)\delta\nu(\vec{r}^\prime-\vec{R}_j)\right\rangle \nonumber \\
&=&\sum_{i=1}^N\left\langle\delta\nu(\vec{r}-\vec{R}_i)\delta\nu(\vec{r}^\prime-\vec{R}_i)\right\rangle\,.
\end{eqnarray}
If all positions $\vec{R}_i$ at the interface are equiprobable,
\begin{equation}
F(\vec{r}, \vec{r}^\prime)=\frac{N}{S}\int_{\cal S} d^2\vec{R}\,\delta\nu(\vec{r}-\vec{R})\delta\nu(\vec{r}^\prime-\vec{R})\,,
\end{equation}
where ${\cal S}$ is the SiO$_2$/HfO$_2$ interface with area $S\to\infty$. As expected, the scattering strength is proportional to the RCS charge density $n_{\rm RCS}=N/S$. We can finally introduce the in-plane coordinate $\vec{r}_\parallel$ and out-of-plane coordinate $x$, and write $F(\vec{r}, \vec{r}^\prime)=NF_{\rm RCS}(\vec{r}_\parallel^\prime-\vec{r}_\parallel, x, x^\prime)$, where:
\begin{equation}
F_{\rm RCS}(\vec{r}_\parallel, x, x^\prime)=\frac{1}{S}\int_{\cal S} d^2\vec{R}_\parallel\,\delta\nu(\vec{R}_\parallel, x)\delta\nu(\vec{R}_\parallel+\vec{r}_\parallel, x^\prime)
\label{eqFRCS}
\end{equation}
is the spatial auto-correlation function of a single RCS charge.

The second-order Green's function $\langle G\rangle(E)$ [Eq. (\ref{eqG2ndorder})] embeds all information about the average density and linear response current in an ensemble of many different realizations of the disorder. We do expect, however, fluctuations to vanish when $S\to\infty$ (thermodynamic limit), so that the probability to find a sample that departs from the average goes to zero. Fig.~\ref{FigvariRCS1} simply shows that the present $W_s=20{\rm\ nm}\times L_s=30{\rm\ nm}$ devices are far too small to kill those fluctuations, so that the number of samples $N_s$ needed to converge the ensemble averaged mobility [Eq.~(\ref{eqmuavg})] is still very large at this scale. We can, unfortunately, neither simulate devices with widths and lengths in the $\mu$m range using NEGF, nor accumulate statistics on hundreds of samples.

We might, therefore, attempt to design a minimal set of test potentials $V_{\rm test}(\vec{r})$ that reproduces the effects of the leading second-order self-energy $\Sigma(\vec{r}, \vec{r}^\prime, E)$. This set of test potentials must satisfy:
\begin{subequations}
\begin{eqnarray}
\langle\delta V_{\rm test}(\vec{r}_\parallel,x)\rangle&=&0 \\
\langle\delta V_{\rm test}(\vec{r}_\parallel,x)\delta V_{\rm test}(\vec{r}_\parallel^\prime,x^\prime)\rangle&=&NF_{\rm RCS}(\vec{r}_\parallel^\prime-\vec{r}_\parallel, x, x^\prime)\,.
\end{eqnarray}
\end{subequations}
Taking the in-plane Fourier transform on both sides,
\begin{subequations}
\begin{eqnarray}
\langle\delta V_{\rm test}(\vec{K}_\parallel, x)\rangle&=&0  \\
\langle\delta V_{\rm test}^*(\vec{K}_\parallel, x)\delta V_{\rm test}(\vec{K}_\parallel^\prime, x^\prime)\rangle&=&NF_{\rm RCS}(\vec{K}_\parallel, x, x^\prime)\delta_{\vec{K}_\parallel,\vec{K}_\parallel^\prime}  \\
&=&N\delta\nu^*(\vec{K}_\parallel, x)\delta\nu(\vec{K}_\parallel, x^\prime)\delta_{\vec{K}_\parallel,\vec{K}_\parallel^\prime}\,, \label{eqVkk}
\end{eqnarray}
\label{eqVtest}
\end{subequations}
where:
\begin{equation}
V_{\rm test}(\vec{K}_\parallel, x)=\frac{1}{S}\int_{\cal S} d^{2}\vec{r}_\parallel\,V_{\rm test}(\vec{r}_\parallel, x)e^{-i\vec{K}_\parallel\cdot\vec{r}_\parallel}\,.
\end{equation}
Note that the diagonal elements $\vec{K}_\parallel=\vec{K}_\parallel^\prime$ in Eq. (\ref{eqVkk}) play a key role in defining the scattering strength. Hence we shall hopefully expedite convergence of the ensemble average by choosing test potentials that satisfy:
\begin{subequations}
\begin{eqnarray}
\langle\delta V_{\rm test}(\vec{K}_\parallel, x)\rangle&=&0 \\
\langle\delta V_{\rm test}^*(\vec{K}_\parallel, x)\delta V_{\rm test}(\vec{K}_\parallel^\prime, x^\prime)\rangle&=&0\text{ if }\vec{K}_\parallel\ne\vec{K}_\parallel^\prime \\
\delta V_{\rm test}^*(\vec{K}_\parallel, x)\delta V_{\rm test}(\vec{K}_\parallel, x^\prime)&=&N\delta\nu^*(\vec{K}_\parallel, x)\delta\nu(\vec{K}_\parallel, x^\prime)\,, \label{eqfitk}
\end{eqnarray}
\label{eqVtest2}
\end{subequations}
The backward Fourier transform of the last equation reads:
\begin{eqnarray}
F_{\rm test}(\vec{r}_\parallel, x, x^\prime)&=&\frac{1}{S}\int_{\cal S} d^2\vec{R}\,\delta V_{\rm test}(\vec{R}, x)\delta V_{\rm test}(\vec{R}+\vec{r}_\parallel, x^\prime) \nonumber \\
&=&NF_{\rm RCS}(\vec{r}_\parallel, x, x^\prime)\,.
\label{eqfitr}
\end{eqnarray}
In other words, the spatial auto-correlation function of each test potential, $F_{\rm test}$, shall match the ensemble averaged auto-covariance function of the disorder. 

As a matter of fact, this strategy is already widely used for surface roughness.\cite{Buran09} The SR samples are generated so that the spatial auto-correlation function, $F(\vec{r}_\parallel)$ [Eq.~(\ref{eqSR2})] matches the ensemble averaged auto-covariance function $F_{\rm SR}(\vec{r}_\parallel)$ [Eq.~(\ref{eqSR})]. In reciprocal space,
\begin{eqnarray}
F_{\rm SR}(\vec{K}_\parallel)&=&\frac{\pi\Delta^2 \ell_c^2}{S\left(1+\vec{K}_\parallel^2 \ell_c^2/2\right)^{3/2}} \nonumber \\
&=&F(\vec{K}_\parallel)=|\delta h(\vec{K}_\parallel)|^2.
\end{eqnarray}
We can therefore choose $\delta h(\vec{K}_\parallel)=e^{i\varphi(\vec{K}_\parallel)}\sqrt{F_{\rm SR}(\vec{K}_\parallel)}$, where $\varphi(\vec{K}_\parallel)$ is a random phase, and transform back to real space to find a suitable SR profile $\delta h(\vec{r}_\parallel)$.

\begin{figure}
\includegraphics[width=.50\columnwidth]{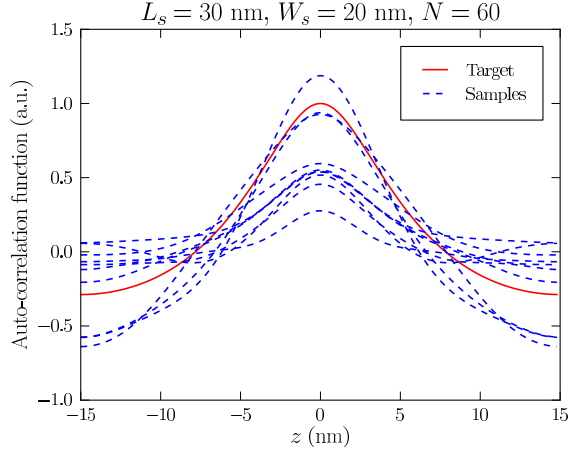} 
\caption{Spatial auto-correlation function $\bar{F}_{\rm test}$ computed along the $z$ axis for 10 test distributions of 60 RCS charges in a $W_s=20{\rm\ nm}\times L_s=30{\rm\ nm}$ sample. The target auto-covariance function $\bar{F}_{\rm target}$ is plotted for comparison. [$t_{\rm Si}=4$ nm, $t_{\rm SiO_2}=1$ nm]\label{FigAFRCS1}}
\end{figure}

How much can the spatial auto-correlation function of a random charge distribution, $F_{\rm test}$, be different from the target $F_{\rm target}=NF_{\rm RCS}$ ? For the purpose of comparisons, we define the $x$-averaged auto-correlation function:
\begin{equation}
\bar{F}(\vec{r}_\parallel)=\frac{1}{L_x^2}\int_{x_0}^{x_1}dx\int_{x_0}^{x_1}dx^\prime\,F(\vec{r}_\parallel, x, x^\prime)\,,
\end{equation}
where $L_x=x_1-x_0$ and the integration range, $[x_0, x_1]$, is typically the upper half of the film (which matters most in the inversion regime). $\bar{F}_{\rm test}$ is plotted for different random, unscreened test distributions on Fig.~\ref{FigAFRCS1}, and compared to $\bar{F}_{\rm target}$. The size of the unit cell is $W_s=20{\rm\ nm}\times L_s=30{\rm\ nm}$, and the density of RCS charges is $n_{\rm RCS}=10^{13}$ cm$^{-2}$ ($N=60$ positive charges). The auto-correlation of the test and target potentials can indeed be very different, which explains the huge variability seen in Fig.~\ref{FigvariRCS1}. This is due to the fact that the size of the sample is not very much larger than the decay length of the auto-correlation function (at variance with the SR profiles discussed in paragraph \ref{subsectionSR}).

A possible set of test potentials that satisfy Eqs. (\ref{eqVtest2}) is given by:
\begin{equation}
\delta V_{\rm test}(\vec{K}_\parallel, x)=\sqrt{N}e^{i\varphi(\vec{K}_\parallel)}\delta\nu(\vec{K}_\parallel, x)\,,
\label{eqRCScont}
\end{equation}
where $\varphi(\vec{K}_\parallel)$ is a random phase shift. However, the backward Fourier transform $\delta V_{\rm test}(\vec{r}_\parallel, x)$ of Eq. (\ref{eqRCScont}) solutions is not, in general, the potential created by a distribution of point charges at the SiO$_2$/HfO$_2$ interface. It might, hence, lead to inappropriate electronic response and electrostatics. In general, there is no distribution of point charges at the SiO$_2$/HfO$_2$ interface that fulfills Eq.~(\ref{eqfitk}) exactly for all $\vec{K}$, $x$ and $x^\prime$. To rank tentative charge distributions, we therefore introduce the least square deviation $\sigma^2$ between $\bar{F}_{\rm test}$ and $\bar{F}_{\rm target}$:
\begin{equation}
\sigma^2=\sum_{\vec{K}_\parallel}\left(\bar{F}_{\rm test}(\vec{K}_\parallel)-\bar{F}_{\rm target}(\vec{K}_\parallel)\right)^2\,,
\end{equation}
where $\bar{F}_{\rm test}(\vec{K}_\parallel)$ and $\bar{F}_{\rm target}(\vec{K}_\parallel)$ are the in-plane Fourier transforms of $\bar{F}_{\rm test}$ and $\bar{F}_{\rm target}$, computed for the unscreened potentials using Fast Fourier Transform on the finite differences grid. We then sample (in parallel) a large number (typically 8192) of random charge distributions, and select the one that minimizes $\sigma^2$.

\begin{figure}
\includegraphics[width=.50\columnwidth]{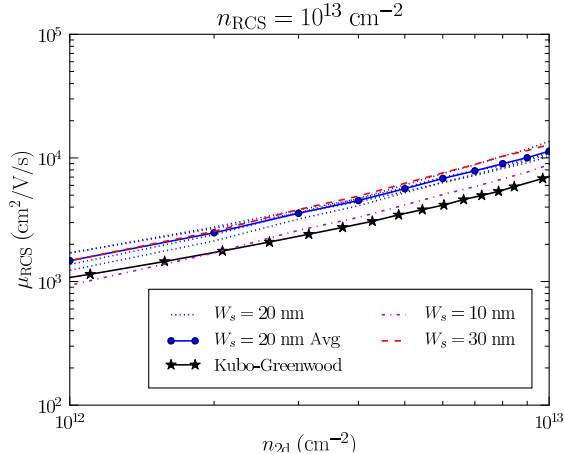} 
\caption{The RCS limited mobility as a function of carrier density, computed in samples with different lengths $L_s$ and widths $W_s$, using optimized test charge distributions at the SiO$_2$/HfO$_2$ interface (positive charges with density $n_{\rm RCS}=10^{13}$ cm$^{-2}$). [$t_{\rm Si}=4$ nm, $t_{\rm SiO_2}=1$ nm]\label{FigvariRCS2}}
\end{figure}

\begin{figure}
\includegraphics[width=.50\columnwidth]{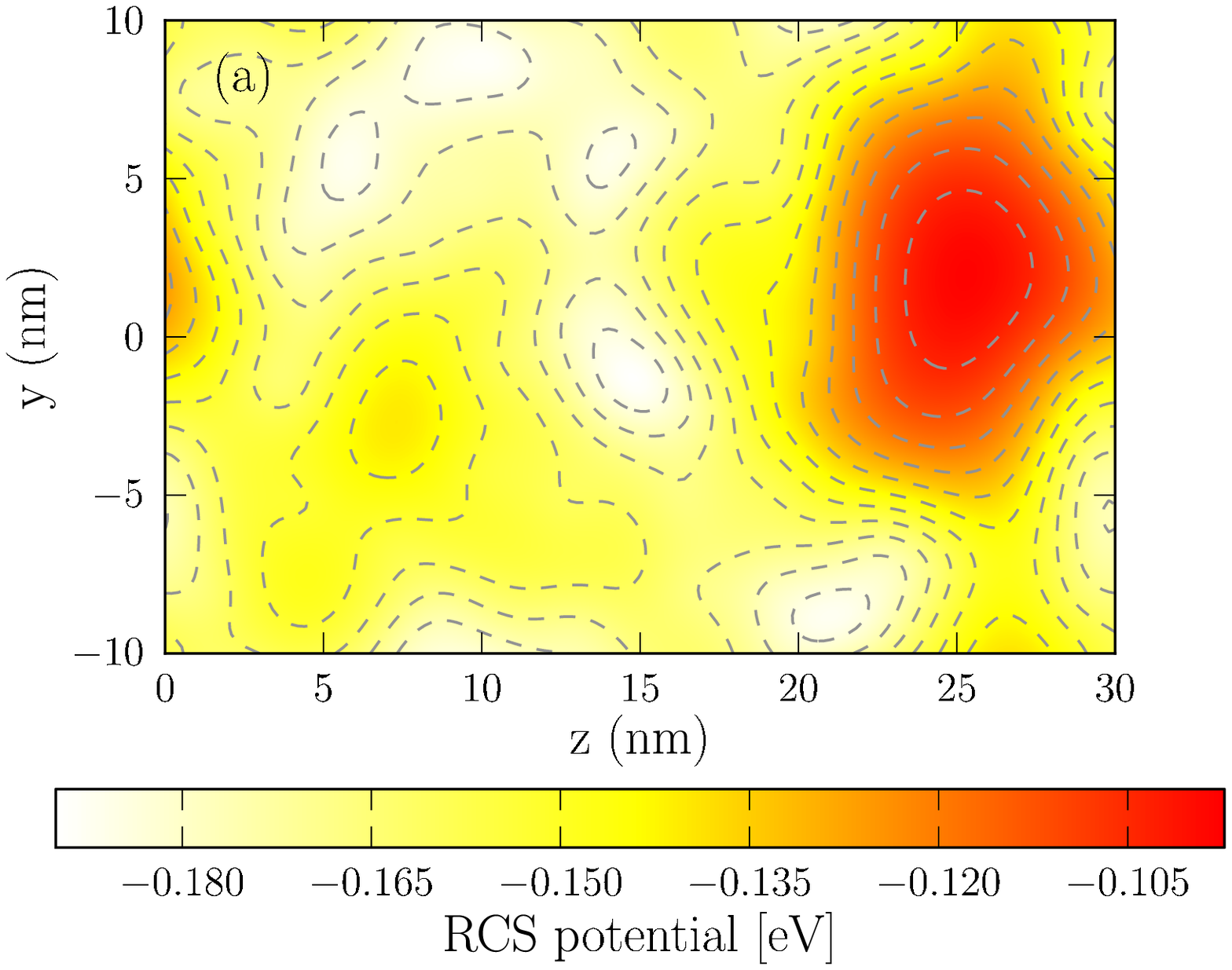} 
\includegraphics[width=.50\columnwidth]{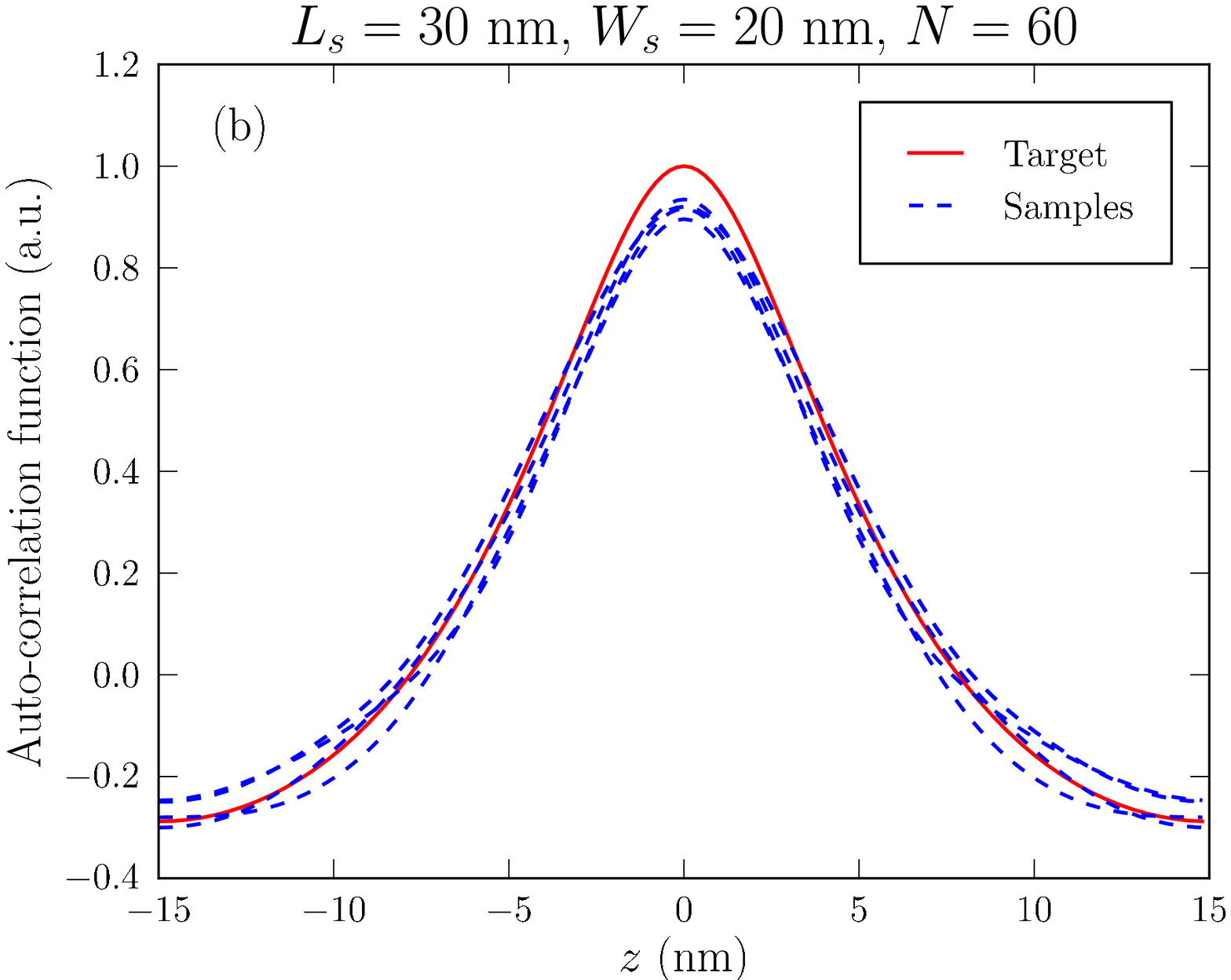} 
\caption{(a) A test RCS potential, plotted 1 nm below the Si/SiO$_2$ interface. (b) Spatial auto-correlation function $\bar{F}_{\rm test}$ computed along the $z$ axis for 4 optimized distributions of 60 RCS charges in a $W_s=20{\rm\ nm}\times L_s=30{\rm\ nm}$ sample. Each optimal distribution was selected out of 8192 configurations, generated with a different random seed. The target auto-correlation function $\bar{F}_{\rm target}$ is plotted for comparison. [$t_{\rm Si}=4$ nm, $t_{\rm SiO_2}=1$ nm]\label{FigAFRCS2}}
\end{figure}

As an example, the mobilities computed using different optimized test charge distributions are plotted in Fig.~\ref{FigvariRCS2} (same film as in Fig.~\ref{FigvariRCS1}). The variability is definitely improved, and the NEGF data are much closer to the KG results (a more detailed comparison will be made in paragraph \ref{subsectioncmpRCS}). This figure also shows the influence of the width $W_s$ of the samples. The sample must be at least 20 nm wide to get reliable results.

The auto-correlation functions of the four $W_s=20$ nm samples of Fig.~\ref{FigvariRCS2} are given in Fig.~\ref{FigAFRCS2}, along with a test RCS potential plotted 1 nm below the Si/SiO$_2$ interface. The target auto-correlation function is, indeed, well reproduced by the four samples.

The above methodology also applies to nanowires and other 1D structures (the $\vec{K}_\parallel$ vector then runs along the nanowire axis). When both surface roughness and RCS come into play, the film or wire is not homogeneous and the potential $\nu(\vec{r}-\vec{R}_i)$ created by a single RCS charge depends on its position $\vec{R}_i$. In that case, we first average $\bar{F}_{\rm target}(\vec{K}_\parallel)$ over (typically 8192) random charge positions at the interface.

To conclude, we have emphasized the importance of having well characterized auto-correlation functions for the disorder in NEGF mobility calculations. This is particularly critical for mechanisms such as RCS, which exhibit long-range correlations.

\section{About Matthiessen's rule}
\label{sectionMatthiessen}

\begin{figure}
\includegraphics[width=.50\columnwidth]{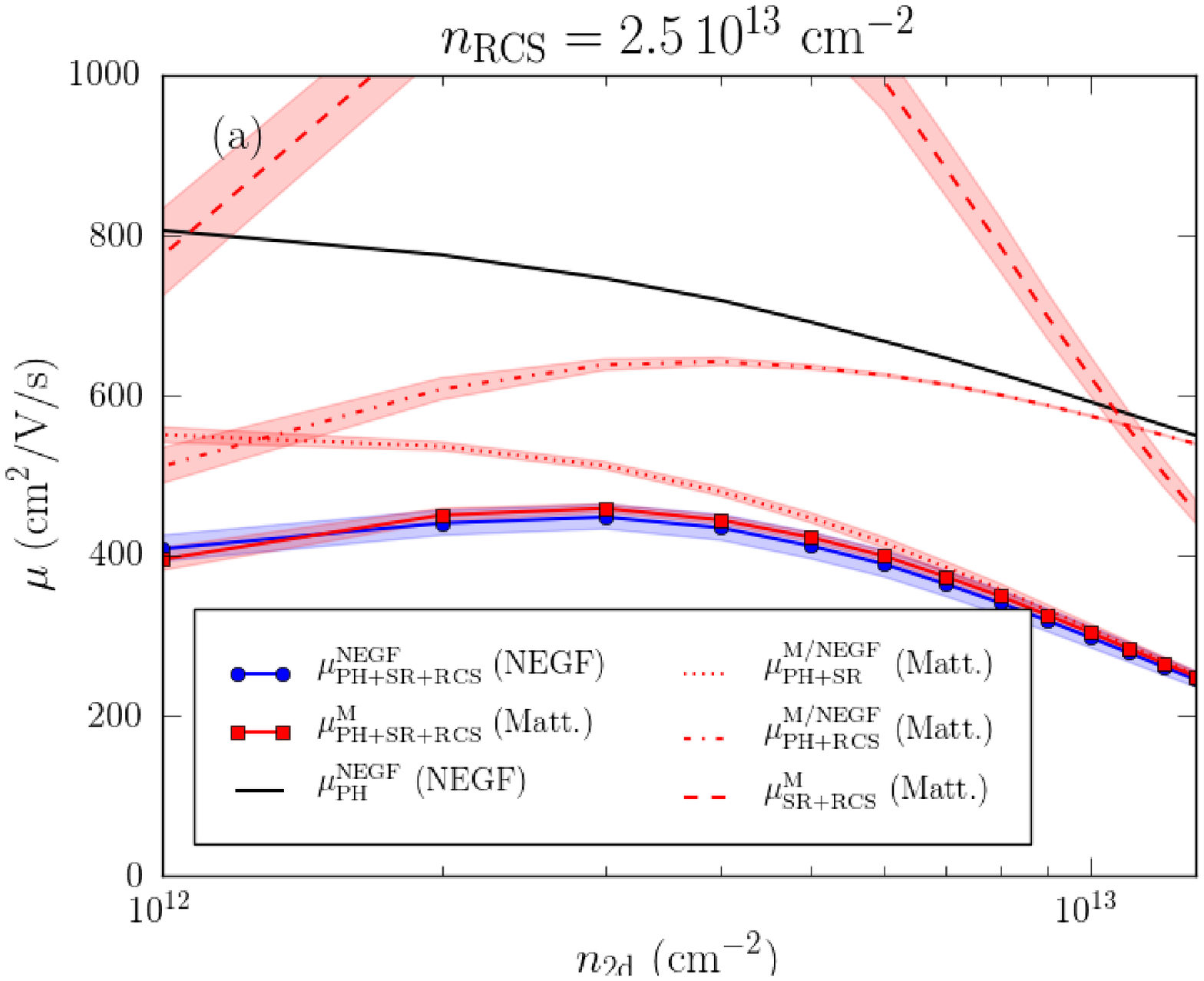} 
\includegraphics[width=.50\columnwidth]{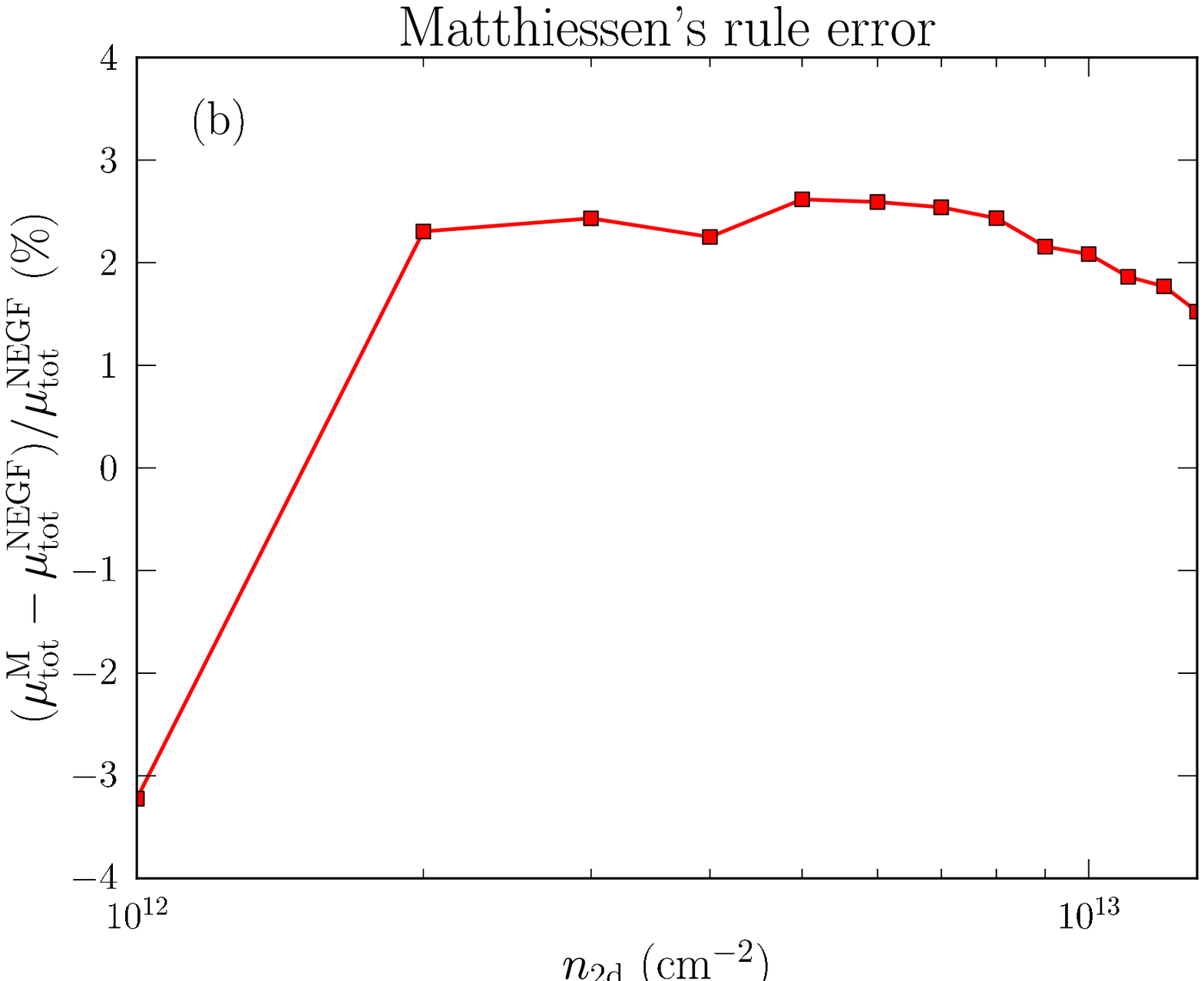} 
\caption{(a) Total NEGF mobility $\mu_{\rm tot}^{\rm NEGF}=\mu_{\rm PH+SR+RCS}^{\rm NEGF}$ including phonons, SR and RCS ($n_{\rm RCS}=2.5\,10^{13}$ cm$^{-2}$), compared with Matthiessen's law $\mu_{\rm tot}^{\rm M}$ on the effective mobilities [Eq. (\ref{eqMatt2})]. Matthiessen's laws on phonons+SR, phonons+RCS and SR+RCS are also plotted for completeness (they are, by design, equivalent to the corresponding NEGF data in the first two cases). The data for $\mu_{\rm SR, eff}$, $\mu_{\rm RCS, eff}$ and $\mu_{\rm tot}^{\rm NEGF}$ were averaged over 4 configurations to reduce the impact of residual variability, shown as a shaded area around each curve. (b) The error made by Matthiessen's law on the total mobility. [$t_{\rm Si}=4$ nm, $t_{\rm SiO_2}=2$ nm]\label{FigMatt}}
\end{figure}

Matthiessen's rule states that the inverse of the total mobility $\mu_{\rm tot}^{\rm M}$ is simply the sum of the inverse of the partial mobilities computed for each different scattering mechanism (e.g. $\mu_{\rm PH}$ for phonons, $\mu_{\rm SR}$ for SR and $\mu_{\rm RCS}$ for RCS):
\begin{equation}
\frac{1}{\mu_{\rm tot}^{\rm M}}=\frac{1}{\mu_{\rm PH}}+\frac{1}{\mu_{\rm SR}}+\frac{1}{\mu_{\rm RCS}}\,.
\label{eqMatt}
\end{equation}
As discussed in section \ref{subsectionPhonons}, the validity of Matthiessen's rule has been questioned in the literature.\cite{Walukiewicz79,Stern80,Takeda81,Saxena85,Fischetti02,Esseni11,Chen13} Indeed, the mobility $\mu_{\rm tot}$ computed with all scattering mechanisms at once can depart significantly from $\mu_{\rm tot}^{\rm M}$. As a matter of fact, Eq.~(\ref{eqMatt}) is exact in a Kubo-Greenwood framework only if {\it i}) the scattering rates of all mechanisms have the same dependence on energy, and {\it ii}) these mechanisms are independent one from each other (no spatial correlations). Although the second condition usually holds for phonons, SR and RCS, the first one does not. Esseni and Driussi\cite{Esseni11} have reported errors $>50\%$ for phonons+Coulomb scattering (at low density) and around $10-15\%$ for phonons+SR (at high density) in bulk MOS transistors.

We argue, however, that the effective SR and RCS mobilities defined by Eq.~(\ref{eqmueff}) follow Matthiessen's rule much better than the usual, single mechanism partial mobilities. This is illustrated in Fig.~\ref{FigMatt}, which compares Matthiessen's rule on the effective mobilities:
\begin{equation}
\frac{1}{\mu_{\rm tot}^{\rm M}}=\frac{1}{\mu_{\rm PH}}+\frac{1}{\mu_{\rm SR, eff}}+\frac{1}{\mu_{\rm RCS, eff}}
\label{eqMatt2}
\end{equation}
with a NEGF calculation including all mechanisms at once. To reduce the impact of residual variability, the data were averaged over 4 different configurations for each single mechanism (SR, RCS) and for the total NEGF mobility $\mu_{\rm tot}$. Mathiessen's holds within 3\% in the whole density range. Similar agreement was obtained on a 7.5 nm thick film, even though the current shows a stronger multi-valley and multi-band character (a condition known to hinder Matthiessen's rule\cite{Fischetti02,Esseni11}), and on a $10\times 10$ nm square nanowire in a trigate configuration.

Of course, one expects Eq.~(\ref{eqMatt2}) to be valid whenever one of the two elastic mechanisms dominates the other. In the high density range for example, where RCS is negligible, $\mu_{\rm SR, eff}$ reproduces, by construction, the total mobility $\mu_{\rm tot}\simeq\mu_{\rm PH+SR}$. However, Fig.~\ref{FigMatt} shows that Matthiessen's rule also holds when SR and RCS have comparable strengths. Inelastic scattering by phonons (included in all calculations) indeed smooths spectral quantities such as the local density of states and the current density, hence reducing the mismatch between different mechanisms that otherwise plagues Matthiessen's rule.

This further motivates the use of Eq.~(\ref{eqmueff}) as a definition of the contribution of a given mechanism to the mobility, and opens the way for a more accurate modeling of the mobility in Technology Computer Aided Design (TCAD) tools.

\section{Comparisons between Kubo-Greenwood and NEGF calculations}
\label{sectionComparisons}

We conclude this paper with a comparison between Kubo-Greenwood and NEGF calculations. We successively discuss phonons, RCS and SR limited mobilities.  

\subsection{Phonons}

\begin{figure}
\includegraphics[width=.50\columnwidth]{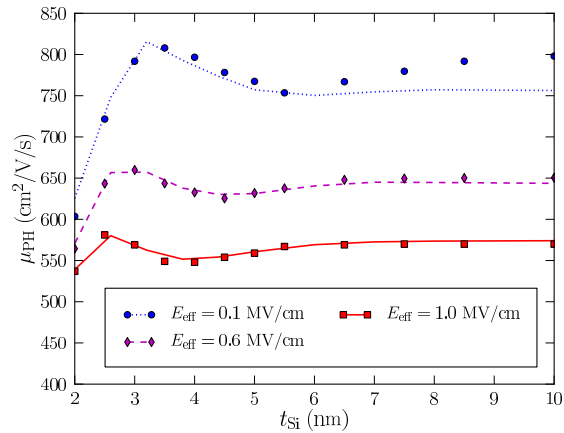} 
\caption{The phonon limited mobility as a function of film thickness, at three effective fields, computed within NEGF (lines) and Kubo-Greenwood (symbols). [$t_{\rm SiO_2}=2$ nm]\label{FigPhonons}}
\end{figure}

KG and NEGF phonon limited mobilities are plotted as a function of Si film thickness $t_{\rm Si}$ in Fig.~\ref{FigPhonons}, for different effective electric fields. They have been computed in the same devices, with the same effective mass model and the same material parameters. As already discussed in Ref.~\onlinecite{Niquet12a}, KG and NEGF are in very good agreement about electron-phonon scattering down to very thin films or wires. In particular, the mobility overshoot around $t_{\rm Si}=3$ nm, which is due to the depletion of the heavy $\Delta^\prime$ valleys off-Gamma into the light $\Delta$ valleys at $\Gamma$, is consistent in both approaches.

\subsection{Remote Coulomb scattering}
\label{subsectioncmpRCS}

\begin{figure}
\includegraphics[width=.50\columnwidth]{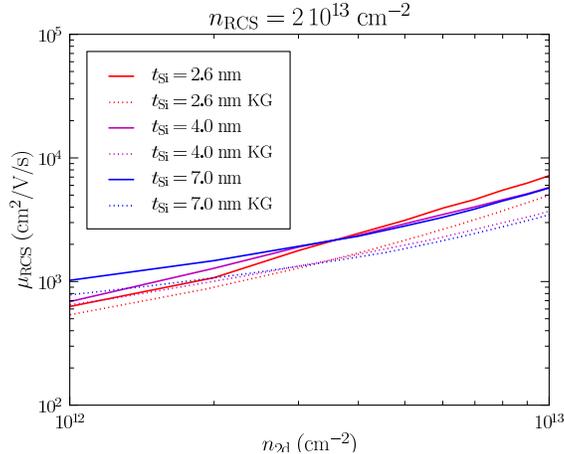} 
\caption{The RCS limited mobility as a function of carrier density, for different $t_{\rm Si}$, computed within NEGF and Kubo-Greenwood. [$t_{\rm SiO_2}=1$ nm]\label{FigRCS}}
\end{figure}

As shown in Fig.~\ref{FigvariRCS2}, the KG model of Sentaurus Device is in reasonable agreement with NEGF. In particular, the slope of the $\mu_{\rm RCS}(n_{\rm 2d})$ curves, which characterizes how the RCS potential is screened, is almost the same in both methods. The NEGF mobility is, however, systematically larger than the KG mobility (whatever the distribution of RCS charges at the interface). The average difference is around $30\%$ over the whole density range. Nonetheless, KG and NEGF show comparable quantitative trends as a function of $t_{\rm Si}$ (and $t_{\rm SiO_2}$), as evidenced in Fig.~\ref{FigRCS}.

Overall, the present KG and NEGF simulations are in line with Ref.~\onlinecite{Toniutti12}, but tend to indicate that the impact of RCS on the mobility is much lower than anticipated from the data of Ref.~\onlinecite{Barraud08}.\cite{NoteBarraud}
 
\subsection{Surface roughness}

\begin{figure}
\includegraphics[width=.50\columnwidth]{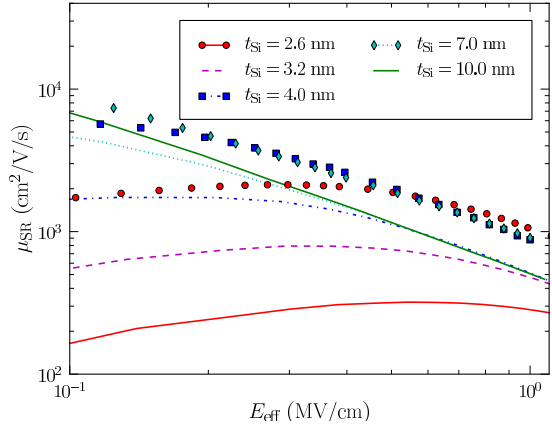} 
\caption{The SR limited mobility as a function of effective field, for different $t_{\rm Si}$, computed within NEGF (lines) and Kubo-Greenwood (symbols). [$t_{\rm SiO_2}=2$ nm, $\Delta=0.47$ nm, $\ell_c=1.3$ nm]\label{FigSR}}
\end{figure}

The NEGF and KG surface roughness limited mobilities are plotted as a function of the effective field in Fig.~\ref{FigSR}, for different film thicknesses $t_{\rm Si}$ ($\Delta=0.47$ nm and $\ell_c=1.3$ nm). As for RCS, the effective KG mobility is computed like the NEGF mobility, from Eq.~(\ref{eqmueff}), but the same conclusions can be reached from a direct KG calculation. It is clear that the NEGF mobility is significantly smaller than the KG mobility, and that the NEGF mobility decreases much faster than the KG mobility when thinning the film. In this respect, the trends followed by the NEGF mobility are much closer to the experimental data of Uchida {\it et al.}.\cite{Uchida01,Uchida03} This strong decrease of the mobility is due to the scattering by film thickness fluctuations.

The disagreement between KG and NEGF might result from {\it i}) the semi-classical nature of Boltzmann transport equation and distribution function; {\it ii}) the second-order perturbation theory (Fermi Golden Rule) behind KG; and {\it iii}) the approximations made in the SR Hamiltonian in KG. We now examine each of these possibilities.

\begin{figure}
\includegraphics[width=.50\columnwidth]{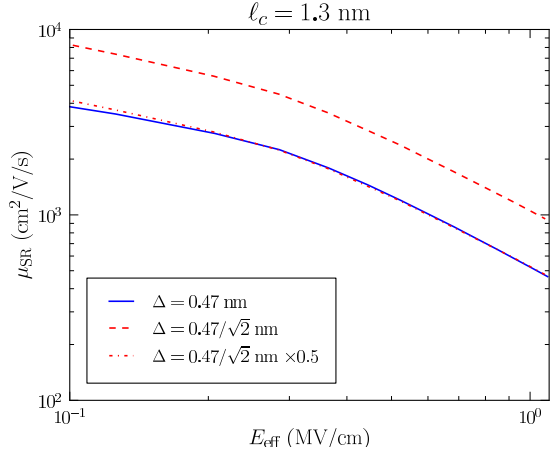} 
\caption{The NEGF SR limited mobility as a function of the effective field, computed for $\Delta=0.47$ nm and $\Delta=0.47/\sqrt{2}=0.33$ nm [$\ell_c=1.3$ nm, $t_{\rm Si}=4$ nm, $t_{\rm SiO_2}=2$ nm]. The mobility shows an almost perfect $\propto 1/\Delta^2$ scaling, as evidenced by the red dashed line.\label{FigscalingSR}}
\end{figure}

The SR, RCS and electron-phonon scattering rates show, admittedly, very different energy and wave vector dependences, but this hardly explains why Boltzmann transport equation reproduces the phonon and RCS limited mobility, but not the SR limited mobility. We therefore presume that the semi-classical nature of this equation is not the primary source of discrepancies with NEGF. As for perturbation theory, Fermi Golden Rule predicts a characteristic, $1/\Delta^2$ behavior for the SR limited mobility, which does not hold at higher order. Interestingly, the NEGF data also show an almost perfect $1/\Delta^2$ scaling around the investigated rms $\Delta=0.47$ nm (see Fig.~\ref{FigscalingSR}). Therefore, Fermi Golden Rule seems to be valid for SR scattering. Hence, approximations on the SR Hamiltonian are most likely responsible for the differences between KG and NEGF. Indeed, SR is explicit in NEGF calculations, and no approximations are made on the SR Hamiltonian and screening (beyond the mean field approximation for Coulomb interactions also made in KG). In KG calculations, SR and screening are ``implicit'' in the sense that they act on a smooth reference device. There are a few approximations in the SR potential (see, e.g., the very detailed derivation of Ref. \onlinecite{Jin09}), whose quantitative accuracy shall be assessed in detail. We suggest that this is the primary target to improve the KG results.

\section{Conclusions}
\label{sectionConclusions}

We have proposed a new method for the calculation of carrier mobilities in a real space NEGF framework that provides accurate results at a reasonable cost. We have also introduced a new paradigm for the definition of the partial mobility associated with a given elastic scattering mechanism, based on Matthiessen's rule with phonons as a reference. We argue that this definition makes better sense in a quantum transport framework, as it mitigates long range interference effects that appear in purely ballistic calculations. As a matter of fact, the partial mobilities obtained that way satisfy Matthiessen's rule for two and three mechanisms much better than the usual, direct single mechanism calculations. We have also emphasized the need for well characterized auto-correlation functions for the disorder in NEGF, in order to mimic the thermodynamic (large samples) limit as best as possible. Finally, we have compared our NEGF mobilities with KG calculations. NEGF and KG are in good agreement for the phonons and RCS, yet not for the SR limited mobilities. We suggest that the weaknesses of KG lie in the approximations made in the treatment of the SR potential and screening, opening the way for further improvements.

We thank S. Barraud for fruitful discussions about RCS. This work was supported by the French National Research Agency (ANR project Quasanova). The NEGF calculations were run at the TGCC/Curie machine using allocations from PRACE and GENCI,\cite{noteCurie} and on the ``Froggy'' platform of the CIMENT infrastructure in Grenoble.\cite{noteFroggy}

\appendix

\section{Details about the NEGF implementation}
\label{AppendixNEGF}

In this appendix, we give details about our implementation of the NEGF equations.

\subsection{NEGF equations and implementation}

The lesser Green's function $G^{v<}(E)$, greater Green's function $G^{v>}(E)$ and retarded Green's function $G^{vr}(E)$ of each valley $v$ embed all information about the charge and current density at energy $E$.\cite{Anantram08} On the finite differences grid, they become matrices that satisfy:
\begin{subequations}
\label{eqGreen}
\begin{eqnarray}
&&\left[E-H^v+eV-\Sigma^{vr}_b(E)-\Sigma^{vr}_s(E)\right]G^{vr}(E)=I \\
&&G^{v\gtrless}(E)=G^{vr}(E)\left[\Sigma^{v\gtrless}_b(E)+\Sigma^{v\gtrless}_s(E)\right]G^{vr}(E)^\dag\,,
\end{eqnarray}
\end{subequations}
where $H$ is the EMA Hamiltonian, $V$ is the electrostatic potential, $\Sigma_b(E)$ is the ``boundary'' self-energy describing the source and drain contacts, and $\Sigma_s(E)$ is the ``scattering'' self-energy accounting for inelastic electron-phonon interactions. The electrostatic potential $V(\vec{r})$ satisfies Poisson's equation:
\begin{equation}
\nabla_\vec{r}\varepsilon(\vec{r})\nabla_\vec{r} V(\vec{r})=-4\pi n(\vec{r})\,,
\end{equation}
where $n(\vec{r})$ is the charge density. The electronic charge $Q_i$ at each point $\vec{R}_i$ of the finite differences grid can be computed from the diagonal elements $G^{v<}_{ii}(E)$ of the lesser Green's function:
\begin{equation}
Q_i=-e\,{\rm Im}\sum_v\int_{-\infty}^{+\infty}\frac{dE}{2\pi}\,G^{v<}_{ii}(E)\,.
\label{eqQG}
\end{equation}
The current in the device can be computed along the same lines.\cite{Anantram08}

Only 1 nm of SiO$_2$ is included in the EMA hamiltonian, on each side of the film. The mesh is homogeneous in this subdomain, with 2\AA\ step, and non homogeneous outside (where only Poisson's equation is solved and the wave functions are assumed to be zero).

Electrons-phonons, surface roughness and remote Coulomb scattering can be included in the calculations. As for phonons, we use the usual diagonal approximation\cite{Jin06} for the self-energy $\Sigma_s$. For intra-valley acoustic phonons,
\begin{equation}
\Sigma^{v\gtrless}_{s,ii}(E)=\frac{1}{\Omega_i}\frac{kT D_{\rm ac}^2}{\rho v_s^2}G^{v\gtrless}_{ii}(E)\,,
\end{equation}
where $kT$ is the thermal energy, $D_{\rm ac}=14.6$ eV is the acoustic deformation potential,\cite{Esseni03a} $\rho=2.33$ g/cm$^3$ is the density of silicon, $v_s=9000$ m/s is the longitudinal sound velocity,\cite{Jacoboni83} and $\Omega_i$ is the elementary volume around point $\vec{R}_i$ of the finite differences grid. For an inter-valleys acoustic or optical phonon,
\begin{equation}
\Sigma^{v\gtrless}_{s,ii}(E)=\frac{1}{\Omega_i}\frac{\hbar (D_{\rm op}^{vv'})^2}{2\rho\omega}\left[\langle N\rangle G^{v'\gtrless}_{ii}(E\pm\hbar\omega)+(\langle N\rangle+1)G^{v'\gtrless}_{ii}(E\mp\hbar\omega)\right]\,,
\end{equation}
where $\hbar\omega$ is the energy of the phonon mode, $\langle N\rangle$ is the average number of phonons in this mode, and $D_{\rm op}^{vv'}$ is a deformation potential. We account for the 3 $f$-type and for the 3 $g$-type inter-valleys processes of Ref.~\onlinecite{Jacoboni83}. We approximate the retarded electron-phonon self-energy $\Sigma^{vr}_s$ as the antihermitic part of $(\Sigma^{v>}_s-\Sigma^{v<}_s)/2$.\cite{Svizhenko03}
 
Born-von-Karman (periodic) boundary conditions are applied in the transverse $y$ direction. The Green's functions can therefore be written:
\begin{equation}
G_{ij}(E)=\sum_{k_y}G_{ij}(k_y, E)e^{ik_y(y_j-y_i)}\,,
\end{equation}
where $k_y$ is the transverse wave vector and $G(k_y, E)$ has the periodicity of the ``supercell'' used for the calculation. The latter is typically $W_s=20$ nm wide, and the first Brillouin zone is sampled with 3 $k_y$ points. While $H$ must be replaced with the Bloch Hamiltonian $H(k_y)$ in Eqs.~(\ref{eqGreen}), the scattering self-energy $\Sigma_s$ remains independent of $k_y$. Dirichlet boundary conditions (constant potential) are applied on the gates, while standard, Neumann boundary conditions (zero normal electric field) are applied along the transport direction.\cite{Ren03}

Equations (\ref{eqGreen}) are solved with a standard Recursive Green's Functions method\cite{Svizhenko02} in a fully coupled mode space approach\cite{Wang04} (192 modes for each valley) on CPUs or graphics cards units (GPU). The latter are highly specialized, parallel units that can process linear algebra operations much faster than traditional CPU cores, thus enabling significant speed-ups.\cite{Nath10,Tomov10} Note that the potential $V$ depends on the charge density in the device, hence on $G^<$, and that the scattering self-energy $\Sigma^<_s$ also depends on $G^<$. One therefore needs to achieve self-consistency on both $V$ and $\Sigma^<_s$. In particular, failure to achieve self-consistency on $\Sigma^<_s$ breaks current conservation (source and drain currents are different).\cite{Baym61,Baym62} It is a common practice to reach self-consistency on $\Sigma^<_s$ for a given $V$ before making any change to the potential,\cite{Luisier09,Cavassilas11} so that the Green's functions used to update $V$ are conserving. However, this strategy, which alternates updates on $\Sigma^<_s$ and $V$, considerably increases the number of iterations needed to achieve global self-consistency (typically $>50$). We find that updating $\Sigma^<_s$ and $V$ at each iteration -- even if the Green's functions used to compute the density are not yet conserving -- expedites convergence to the same fixed point in only 5 to 25 iterations depending on the bias conditions. The Green's function $G^<$ computed at a given iteration is directly used as input for the self-energy $\Sigma^<_s$ of the next iteration, while the variations of the potential $V$ are damped with a Newton-Raphson-like correction (see below).\cite{Ren03} The modes are also updated at each iteration to account for the changes in the potential. The convergence criteria are {\it i}) variations of the charge density $<0.01\%$ for three consecutive iterations, {\it ii}) variations of the average current $<0.1\%$, and {\it iii}) current conserved within $1\%$ along the device. Integrations such as Eq. (\ref{eqQG}) are performed on a regular grid of 256 energy points extending from $\mu_s-0.2$ eV to $\mu_s+16kT$, where $\mu_s$ is the chemical potential of the source. With these parameters, we estimate the error on the current to be $<0.25\%$.

The code is parallelized over the loops on $k_y$ points and energies.

\subsection{Newton-Raphson-like correction to the potential}

Poisson's equation can formally be written:\cite{Ren03}
\begin{equation}
\nabla_\vec{r}\varepsilon(\vec{r})\nabla_\vec{r} V(\vec{r})=-4\pi n[V](\vec{r})\,,
\label{eqPoisson}
\end{equation}
where we have emphasized that $n(\vec{r})$ is a functional of the potential $V(\vec{r})$. This non-linear problem can in principle be solved with the Newton-Raphson method. Starting from an arbitrary potential $V_0(\vec{r})$, we look for a correction $\delta V(\vec{r})=V(\vec{r})-V_0(\vec{r})$ such that:
\begin{equation}
\nabla_\vec{r}\varepsilon(\vec{r})\nabla_\vec{r}\delta V(\vec{r})=-4\pi n[V_0+\delta V](\vec{r})-\nabla_\vec{r}\varepsilon(\vec{r})\nabla_\vec{r} V_0(\vec{r})\,.
\label{eqPoissonNR}
\end{equation}
We then linearize the right-hand side:
\begin{equation}
n[V_0+\delta V](\vec{r})=n[V_0](\vec{r})+\int d^3\vec{r}^\prime D[V_0](\vec{r},\vec{r}^\prime)\delta V(\vec{r}^\prime)\,,
\label{eqnlin}
\end{equation}
where $D[V](\vec{r},\vec{r}^\prime)=\delta n[V](\vec{r})/\delta V(\vec{r}^\prime)$ is the functional derivative of $n[V](\vec{r})$ with respect to $V(\vec{r}^\prime)$. The resulting equation can be solved on the finite differences grid with standard linear algebra routines. We next iterate from the new solution $V_1(\vec{r})=V_0(\vec{r})+\delta V(\vec{r})$, until convergence. The Newton-Raphson method converges in principle much faster than straightforward fixed-point iteration (that is, solving $\nabla_\vec{r}\varepsilon(\vec{r})\nabla_\vec{r} V_1(\vec{r})=-4\pi n[V_0](\vec{r})$ and iterating until self-consistency).

We do not know, however, the explicit form of the functional $n[V]$ and of its derivatives. Yet we might design approximations for $D[V](\vec{r},\vec{r}^\prime)$, solve Eqs.~(\ref{eqPoissonNR}) and (\ref{eqnlin}) for the input potential $V_m(\vec{r})$ and the output density $n[V_m](\vec{r})$ of the $m^{\rm th}$ NEGF iteration, and use the resulting $V_{m+1}(\vec{r})=V_m(\vec{r})+\delta V(\vec{r})$ as input for the $(m+1)^{\rm th}$ iteration. Since $V_{m+1}(\vec{r})$ anticipates over the response of the density through the $D[V](\vec{r},\vec{r}^\prime)$ kernel, we shall hopefully achieve self-consistency much faster.

Practically, we make a local density approximation for $D(\vec{r},\vec{r}^\prime)$:
\begin{equation}
D(\vec{r},\vec{r}^\prime)=e\frac{dn_0(\vec{r}, \mu(\vec{r}))}{d\mu}\delta(\vec{r}-\vec{r}^\prime)\,,
\label{eqDnr}
\end{equation}
where:
\begin{equation}
n_0(\vec{r}, \mu)=-e\int dE\,\rho_0(\vec{r}, E)f_{\rm FD}(E-\mu)\,,
\end{equation}
$\rho_0(\vec{r}, E)$ is the local density of states and $f_{\rm FD}$ is the Fermi-Dirac distribution function. In Eq.~(\ref{eqDnr}), the local chemical potential $\mu(\vec{r})$ is computed so that $n_0(\vec{r}, \mu(\vec{r}))$ matches $n[V_m](\vec{r})$ at each point $\vec{r}$. We further approximate $\rho_0(\vec{r}, E)$ as  $\rho_s(x, y, E)$, the local density of states in the source, in the absence of external potential, computed once for all at the beginning.\cite{noterhos} This approximation for $D(\vec{r},\vec{r}^\prime)$ takes quantum confinement into account and is therefore more accurate than a Fermi integral formula for $n_0$.\cite{Ren03} It makes an excellent preconditioner for the self-consistent Poisson iteration, even far out of equilibrium.\cite{Nguyen13}

\bibliography{./NEGF}

\begin{thebibliography}{10}
\bibitem{NoteBarraud}%
  \BibitemOpen
  \bibinfo {note} {There is possibly a factor $1/(4\pi^2)$ missing in Eq. (35)
  [hence a factor $1/(2\pi)$ missing in Eq. (36)], and a factor $g_s=2$
  accounting for spin degeneracy missing in Eq. (38), which would result in an
  underestimation of the RCS mobility by a factor $4\pi$ in
  Ref.~\onlinecite{Barraud08}.}\BibitemShut{Stop}%
\bibitem{noterhos}%
  \BibitemOpen
  \bibinfo {note} {$\rho_s(E)=-{\rm Im\ } G_s(E+i\eta)/\pi$ is computed from
  the retarded bulk Green's function $G_s$ of the source, a by-product of
  standard decimation routines for the contact self-energy. A large imaginary
  part $\eta$ is added to the energy to smooth out rapid variations of the
  density of states.}\BibitemShut{Stop}%
\end{thebibliography}

\end{document}